\documentclass[11pt]{article}
\usepackage{amssymb}
\usepackage{epsf}
\usepackage{lscape}
\usepackage[utf8]{inputenc}
\usepackage{amssymb}
\usepackage{epsf}
\usepackage{lscape}
\usepackage[english]{babel}
\begin{document}
\begin{center}
{\bf  Asymptotics of three-body bound state radial wave functions
of halo nuclei involving  two charged particles\\
\vspace*{1cm}}
{\bf R. Yarmukhamedov \\}
\vspace*{3mm}
 E-mail: rakhim@inp.uz  \\
\vspace*{9mm} { \it Institute of Nuclear Physics,  Tashkent
100214, Uzbekistan }  
\end{center}
\vspace*{5mm}
\begin{abstract}
Asymptotic expressions for the radial  and full  wave functions  of
a three--body bound halo  nuclear system with two charged particles
in relative coordinates are obtained in explicit form, when the
relative distance between two particles tends to infinity.   The
obtained asymptotic forms are applied to the analysis of the
asymptotic behavior of  the   three-body ($pn\alpha$) wave functions  for
the halo ($E^*$=3.562 MeV, J$^\pi$=0$^+$, T=1) state of ${\rm{^6Li}}$ derived by D. Baye within the Lagrange-mesh method for two forms of the $\alpha N$ -potential. The agreement between the calculated
wave function and the asymptotic formula is excellent  for 
distances up to  30 fm. Information  about the values
of the three-body asymptotic normalization functions is extracted. It is shown that
the extracted values of the three-body asymptotic  normalization
function are sensitive to the form of the   $\alpha N$-potential. The mirror symmetry is revealed for the three-body asymptotic  normalization functions derived for the isobaric (${\rm{^6He}},{\rm{^6Li}}^*$) pair. 
\end{abstract}

\section{INTRODUCTION}

\hspace{0.7cm} Study of    structure of   light exotic nuclei, lying
near the drip lines and   so-called halo nuclei is   one of the most
interesting topics of low-energy nuclear physics
[1--13]. It has revealed a number of features inherent
only for these nuclei but not for normal (non-halo) nuclei, such as
rather low separation energies of the external
(``valence'') nucleons,  large radii and narrow peaks
observed in the breakup probability distribution. Two-nucleon halo
nuclei are particularly striking since the lowest breakup channel
is a three-body (core+``valence'' nucleons) channel
because of the fact that their two-body subsystems are unbound.
Therefore, for these nuclei a main term of the asymptotics of   wave
functions  must be determined by proper three-body asymptotics
\cite{Mer1}.

For more than ten years several works have been devoted to the study
of the   asymptotics of  three-body bound state radial wave
functions of   halo nuclei [15-19]. In Refs.\cite{Blokh99,Yar2009}, the asymptotic expressions have been derived for
three-body systems with two neutrons for the case of short-range (nuclear) interactions. These asymptotics  were derived  in the context of hyperspherical coordinates for large values of the hyperradius $R$
($R\to\infty$), which means that either both of the Jacobi coordinates tend
to infinity \cite{Blokh05} or one of them tends to infinity and another of them is very small \cite{Yar2009}.  In \cite{Blokh05}, the result of \cite{Blokh99}
was generalized for three-body systems including    two charged particles  with taking into account    Coulomb-nuclear interactions.   The obtained asymptotic expressions
contain an  exponential function depending on the   hyperradius $R$
\cite{Mer1} but also involve a factor that can influence noticeably
the asymptotic values of the three-body wave function for some
directions   in the configuration space determined by the hyperangle
$\varphi$, where $\varphi=arctan(y/x)$ ($x$ and $y$ are a pair of
modified Jacobi coordinates \cite{Mer1}). In Refs.\cite{Blokh99,Yar2009} and
\cite{Blokh05}, these asymptotic expressions have been compared with
the asymptotic behaviour  of three-body ($nn\alpha$ and
$\alpha\alpha n$) radial wave functions of the $^6$He and $^9$Be
nuclei respectively derived in Refs.\cite{Kuk95} and \cite{Voron95}
within the framework of the multicluster stochastic dynamical   model, respectively,  where only two forms for the $n\alpha$ and $\alpha\alpha$ potentials are used.
 In Refs. [15--17], information was
obtained about the three-body asymptotic normalization function
(TBANF) as a function of the hyperangle $\varphi$.  However,  
as  is   revealed in [15--17],   the three-body   radial wave functions  of the $^6$He and $^9$Be nuclei \cite{Kuk95,Voron95} have a correct
asymptotic behaviour in an asymptotical region within the interval
with narrow width. Besides, the   binding energies of   $^6$He and
$^9$Be nuclei  in the ($\alpha+2n$)-and ($2\alpha+n$)-channels
calculated in \cite{Kuk95} and \cite{Voron95}, respectively, differ  noticeably
from the experimental ones and, consequently, this circumstance may
also noticeably influence  the TBANFs [15--17].

In Ref.\cite{Yar02}, the asymptotic expression has been derived for
three-body radial wave functions of the  halo nucleus with two
valence neutrons in the context of the relative core-neutron
coordinates for large values of each of them. It was revealed that
the  asymptotic expression obtained in   \cite{Yar02} is not
directly comparable to that derived in   \cite{Blokh99,Yar2009} but
it becomes equivalent to the asymptotic form derived in  
\cite{Blokh99,Yar2009} when the   core is heavy.  As a result, in
Ref.\cite{Yar02}, information  about the values of the   TBANFs has
been obtained by means of comparative analysis of the obtained
asymptotic forms with the asymptotic behavior of the corresponding
model three-body ($ nn\alpha$) wave functions for  the  $^6$He nucleus
derived within the Lagrange-mesh technique \cite{Bay94,Bay97} by
using the $\alpha n$ potential taken from \cite{Kan79}. There it was
shown that the Lagrange-mesh approximate wave function for the bound
$^6$He state is in good agreement with the asymptotic expression
over larger values of the relative core-neutron coordinates (up to
20 fm) and information was obtained about the TBANF as a function of
the   ratio of the relative core-neutron coordinates. It should be
noted  that in \cite{Bay94,Bay97} the   binding energy calculated
for the ground state of $^6$He is in excellent agreement with the
experimental one for the employed $nn$ and $n\alpha$ potentials.

In this connection it should  be emphasized that the TBANF is a
fundamental characteristic of the three-body bound system, which
plays the same role as the asymptotic normalization coefficient of
the radial wave function for the two-body bound system
\cite{Blokh77,Loch78}. Consequently, the TBANF is determined by the
dynamics of strong interactions and, so,   carries information about
two-particle (nuclear) interactions in the three-body bound system.
For example, as is shown in  \cite{Blokh99,Yar2009} and
\cite{Blokh05}, the extracted values of the TBANFs for the $^6$He
and $^9$Be nuclei are highly sensitive to the forms of the $\alpha
n$ and $\alpha\alpha$ potentials, respectively. Consequently,
knowledge of the TBANF, which plays an important role in nuclear
structure, allows one to get the information both on the three-body
structure of halo nuclei and on types of two-particle
(cluster-cluster, cluster-nucleon and nucleon-nucleon) interactions.
Besides, as is shown in \cite{Blokh2013}, the anomalous  asymptotics of  radial overlap integrals for bound systems $a$ of four bodies  in the ($b$+$c$)-channel is expressed in terms of  the TBANF.
Therefore, systematic collection of data about TBANFs for different
halo nuclei must be extremely encouraged now.

In the present work, the asymptotic behavior of the radial wave
functions of a three-body bound (123)-system involving  two charged
particles is studied  for the large relative coordinates $r_{13}$
and $r_{23}$. The results are compared with the   Lagrange-mesh
approximate three-body ($pn\alpha$) radial wave functions derived by D. Baye \cite{Baye2007}
for the    halo ($E^*$=3.562 MeV;
$J^\pi=0^+;T=1$) state of ${\rm{^6Li}}$, which  is the isobar--analog one for the
ground state of ${\rm{^6He}}$ for which $J^\pi=0^+$ and $T=1$ also.

The content of this paper is as follows. In Section 2, the asymptotic expression for the radial component of three-body wave function is derived.  In Section 3, the asymptotic formula is used for testing the asymptotic behavior of the partial waves of
the three-body ($pn\alpha$) wave function for   the    halo ($E^*$=3.562 MeV)   state of the $^6$Li nucleus   by using two kinds of the $\alpha N$ potential, which  
 is briefly described in Appendix B.   In Section 3,  the information about the TBANFs is also analysed and discussed. Conclusion are given in Section 4 and the asymptotic formula for the total three-body wave function is derived in Appendix A.

\section{Asymptotic behavior of three-body radial wave functions with two charged particles}
\vspace{0.5cm}

\hspace{0.2cm} One considers the bound three-body system (123) consisting of two ``valence'' nucleons and
two charged particles (say, particle 1 and  a core 3).
 Let us write ${\bf r}_{ij}={\bf r}_i-{\bf r}_j$ for the relative radius
 vector and   ${\bf q}_{ij}$ for
corresponding relative momentum, where ${\bf r}_k$ is a radius vector of
 the center of mass of the particle $k$.  If
$m_j$ is the mass of particle $j$, we denote
$\mu_{ij}=m_im_j/m_{ij}$ and $\mu_{(ij)k}=m_{ij}m_k/m$ the reduced
masses of the $ij$ and $(ij)k$ systems, respectively, where
$m_{ij}=m_i+m_j$ and $m=m_1+m_2+m_3$. The  $^6$Li($pn\alpha$) nucleus
  considered within a three-body model can
be used  as an example.

The Fourier transformation for  the radial  three-body wave functions $\Psi_{\nu}(r_{23},r_{{13}})$   can be presented   to the form
$$
\Psi_{\nu}(r_{23},r_{{13}})=\frac{(-1)^{l_{23}+l_{{13}}+1}}{(2\pi)^4}
\frac{1}{r_{23}r_{{13}}}
$$
\begin{equation}
\times \int_{-\infty}^{\infty} dq_{23} q_{23} e^{iq_{23}r_{23}}
f_{l_{23}}(q_{23}r_{23})\int_{-\infty}^{\infty} dq_{{13}} q_{{13}}
e^{iq_{{13}}r_{{13}}}f_{l_{{13}}}(q_{{13}}r_{{13}})\Psi_{\nu}(q_{23},q_{{13}}),
\label{8}
\end{equation}
which is derived from Eqs. (6) and (15) of Ref. \cite{Yar02}.
Herein  $\Psi_{\nu}(q_{23},q_{{13}})$ is the partial three-body wave functions in momentum space, which can be  determined from  the relation 
\begin{equation} \Psi_{\nu}(q_{23},q_{{13}})=(4\pi)^{-1}\int d\Omega_{{\mathbf
q}_{23}} d\Omega_{{\mathbf q}_{13}}
Y^*_{l_{23}l_{{13}}LM_L}(\hat{{\mathbf q}}_{23},\hat{{\mathbf
q}}_{13}) \Psi({\mathbf q}_{23},{\mathbf q}_{13}),\label{3}
\end{equation}
 and
\begin{equation}
f_l(x)=\sum_{n=0}^l\frac{(l+n)!}{n!(l-n)!}
\frac{1}{(-2ix)^n}, \label{10}
\end{equation}
where $\Psi({\mathbf q}_{23},{\mathbf q}_{13})$ is the total three-body wave function in the momentum representation,  $\nu=\{l_{23}l_{{13}}Ls_{12}S\}$, $ l_{ij}$ is the relative orbital momentum of particles $i$ and $j$; $ {\mathbf L}= {\mathbf  l}_{23}+
 {\mathbf  l}_{13}$, ${\mathbf s}_{12}={\mathbf s}_1+{\mathbf s}_2$ and ${\mathbf S}={\mathbf s}_{12}+{\mathbf s}_3 $;  ${\mathbf s}_j$ is the
spin of particle $j$ and $Y^*_{l_{23}l_{{13}}LM_L}(\hat{{\mathbf q}}_{23},\hat{{\mathbf q}}_{13})$ is the eigenstates  of the square of the
total orbital angular momentum ${\mathbf L}$ of the three-body (123) system
and of its projection $L_z$ along the z axis, which has the form  
\begin{equation}
 Y_{l_{23}l_{13}LM_L}(\hat{{\mathbf q}}_{23},\hat{{\mathbf
q}}_{13})= \sum_{\nu_{23}\nu_{13}}
C_{l_{23}\nu_{23}l_{{13}}\nu_{{13}}}^{LM_L}
Y_{l_{23}\nu_{23}}(\hat{{\mathbf q}}_{23})
Y_{l_{13}\nu_{13}}(\hat{{\mathbf q}}_{13})\label{2}
\end{equation}
in which $C_{a\alpha\, b\beta}^{c\gamma}$ is a Clebsh-Gordan
coefficient and $\hat{{\mathbf q}}_{ij} = {\mathbf q}_{ij}/q_{ij}$.
 
  The momentum-space wave function $\Psi({\mathbf q}_{23},{\mathbf
q}_{13})$ is related to the   vertex function $W({\mathbf
q}_{23},{\mathbf q}_{13})$ for the virtual decay
\begin{equation}
(123) \rightarrow 1+2+3\label{11}
\end{equation}
by the relation \cite{Bl2}
\begin{equation} \Psi({\mathbf q}_{23},{\mathbf q}_{{13}})=-\frac{W({\mathbf
q}_{23},{\mathbf q}_{{13}})} {\varepsilon+\varepsilon_a}.\label{12}
\end{equation}
Herein the energy $\varepsilon$ reads \cite{Bay94}
\begin{equation} \varepsilon
= \frac{{\mathbf q}_{{13}}^2}{2\mu_{(23)1}} + \frac{({\mathbf
q}_{23}+\lambda_2{\mathbf q}_{{13}})^2} {2\mu_{23}} \label{13}
\end{equation}
\begin{equation} =
\frac{{\mathbf q}_{23}^2}{2\mu_{({31})2}} + \frac{({\mathbf
q}_{13}+\lambda_1{\mathbf q}_{23})^2} {2\mu_{{31}}},\label{14}
\end{equation}
 where  $\lambda_1=m_1/m_{{31}}$ and
$\lambda_2=m_2/m_{23}$, and $\varepsilon_a$ is the binding energy of the system
 $a\equiv (123)$ in respect to the virtual decay (\ref{11})  $ (\varepsilon_a >0)$.
We use $\hbar=c=1$ throughout.

According to Ref.\cite{Mer1}, if, in the bound (123)-system,
two-particle subsystem ($ij$) ($ij$= 12,23,31) can be bound with the
binding energy $\varepsilon_{ij}$ with respect to the
($i+j$)-channel, the vertex function $W({\mathbf q}_{23},{\mathbf
q}_{13})$ has so-called two-particle singularities as a function of
the relative kinetic energy of the particles $i$ and $j$, $E_{ij}$,
at $E_{ij}=-\varepsilon_{ij}$.
 For the two-particle (12)-and (23)-subsystems,
where the Coulomb interactions are absent, these two-particle
singularities are  pole ones, and for the two-particle
({13})-subsystem the two-particle singularity is a power branch
point arising due to
 the Coulomb interaction \cite{Dol}. Besides, according to
Refs.\cite{Mer1,Mukh}, the vertex function $W({\mathbf q}_{23},{\mathbf q}_{13})$
has the three-body branch point singularity at $\varepsilon=-\varepsilon_a$.
 As   seen from Eq.(\ref{12}),
 the wave function $\Psi ({\mathbf q}_{23},{\mathbf q}_{13})$ possesses the same
 singularities  as the vertex function $W({\mathbf q}_{23},{\mathbf q}_{13})$.

In the general case the asymptotic form of the radial wave function
$\Psi_{\nu}(r_{23},r_{{13}})$ at $r_{23}\to\infty$ (or $r_{{13}}\to\infty$) is
determined by both the two-body cluster singularities, which  are associated
 with a formation of possible bound states in the two-particle subsystems
\cite{Mer1}, and by the three-body singularity at $\varepsilon=-\varepsilon_a$
of the  partial wave functions in the momentum space
$\Psi_{\nu}(q_{23},q_{{13}})$.
 Explicit form of the cluster  asymptotic can easily be derived by
using the results of Ref.\cite{Mer1}. But, as a rule, two-body
$(ij)$ subsystem in bound three-body (123) halo nuclei are unbound, 
for example, the $^6$Li($pn\alpha$) nucleus  in the second excited ($E^*$=3.562 MeV)
state being the isobar-analog  to the ground state of $^6$He($nn\alpha$).  For
such nuclei an asymptotic form of the three-body wave function is
determined by the proper three-body asymptotics of function
$\Psi_{\nu}(r_{23},r_{{13}})$ at $r_{23}\to\infty$ (or
$r_{13}\to\infty$). Therefore, here we are interested in the
asymptotic expression for $\Psi_{\nu}(r_{23},r_{{13}})$ at
$r_{23}\to\infty$ (or $r_{13}\to\infty$) determined by means of the
extraction of a contribution from the  aforesaid three-body branch
point singularity into the radial wave function
$\Psi_{\nu}(r_{23},r_{{13}})$ given by Eqs.(\ref{8}), (\ref{3}) and
(\ref{12}).  To this end, one can make use in Eq. (\ref{12})
  the following singular part at
$\varepsilon\to-\varepsilon_a$, $W^{(s)}({\mathbf
q}_{23},{\mathbf q}_{13})$, of the vertex function $W({\mathbf
q}_{23},{\mathbf q}_{13})$ \cite{Blokh05,Mukh}:
 \begin{equation}
W({\mathbf q}_{23},{\mathbf q}_{13})\simeq W^{(s)}({\mathbf
q}_{23},{\mathbf q}_{13}) =\Gamma(1-\eta_a(q_{13}))\tilde W({\mathbf
q}_{23},{\mathbf q}_{13})
\left(\frac{\varepsilon+\varepsilon_a}{4\varepsilon_a}\right)^{\eta_a(q_{{13}})},
\,\,\,
 \varepsilon\to-\varepsilon_a,\label{15} 
 \end{equation}
where $\eta_a(q_{{13}})=iz_1z_3e^2 \mu_{13}/q_{13},\,\, z_je$ is a
charge of the particle $j$,  and the function $\tilde W({\mathbf
q}_{23},{\mathbf q}_{13})$  at the point
$\varepsilon=-\varepsilon_a$ is   regular   part of the vertex function
 $W({\mathbf q}_{23},{\mathbf q}_{13})$, which coincides with the on-shell vertex
function (OSVF) for the virtual decay (\ref{11}) \cite{Mukh}, i.e.,
with a value of the vertex function $W({\mathbf q}_{23},{\mathbf
q}_{13})$ in which its arguments  satisfy the relation
${\mathbf q}_{23}^2/2\mu_{(13)2}+({\mathbf q}_{13}+ \lambda_1{\mathbf
q}_{23})^2/2\mu_{13}=-\varepsilon_a$.

Taking into account the expressions (\ref{12}) and (\ref{13}), one
can perform
 the integration over the angular variables
$\Omega_{{\mathbf q}_{23}}$ and $\Omega_{{\mathbf q}_{13}}$ in
Eq.(\ref{3}). To this end, in Eq.(\ref{15}), one makes use of the partial wave
expansion for the OSVF:
 \begin{equation}
 {\tilde {W}}({\mathbf q}_{23},{\mathbf q}_{13}) = 4\pi \sum_{l_{23}'l_{13}'L'M_{L'}}
Y_{l_{23}'l_{13}'L'M_{L'}}(\hat {{\mathbf q}}_{23},\hat {{\mathbf q}}_{13})
 W_{l_{23}'l_{13}'L's_{12}S}(q_{23},q_{13}),\label{16}
\end{equation}
where $ W_{l_{23}'l_{13}'L's_{12}S}(q_{23},q_{13})$ is a partial OSVF, and of
the following presentation for the function
 $[(\varepsilon+\varepsilon_a)/4\varepsilon_a]^{\eta_a(q_{{13}})}/
(\varepsilon+\varepsilon_a)$:
\begin{equation}
\frac{1}{\varepsilon_a+\varepsilon}\left (\frac{\varepsilon+\varepsilon_a}
 {4\varepsilon_a}\right )^{\eta_a(q_{{13}})}=4\pi\sum_{L''M_{L''}}A_{L'';\eta_a(q_{{13}})}
(q_{23},q_{{13}})Y^*_{L''M_{L''}}(\hat{{\mathbf q}}_{23})
 Y_{L''M_{L''}}(\hat {{\mathbf q}}_{{13}}) .\label{17}
\end{equation}
  Herein  the function $A_{L'';\eta_a(q_{{13}})}(q_{23},q_{{13}})$ is determined by the
 relation
\begin{equation}
 A_{L'';\eta_a(q_{{13}})}(q_{23},q_{{13}})=\frac{1}{2}\int_{-1}^1 P_{L''}(z)
\left(\frac{\varepsilon
 +\varepsilon_a}{4\varepsilon_a}\right)^{\eta_a(q_{{13}})}\frac{dz}{\varepsilon_a
 +\varepsilon}\label{18}
\end{equation}
\begin{equation}
 = \frac{(-1)^{L''}}{4\varepsilon_a}
\left(\frac{4m_3\varepsilon_a}{q_{23}q_{{13}}}\right)^{1-\eta_a(q_{{13}})}
\frac{e^{i\pi\eta_a(q_{{13}})}} {\Gamma(1-\eta_a(q_{{13}}))}
[\zeta^2(q_{23},q_{{13}})-1]^{\eta_a(q_{{13}})/2}Q_{L''}^{-\eta_a(q_{{13}})}[\zeta(q_{23},q_{{13}})],
\label{19}
\end{equation}
where
\begin{equation}
\varepsilon+\varepsilon_a=\frac{q_{23}q_{{13}}}{m_3}[\zeta(q_{23},q_{{13}})+z],\label{20}
\end{equation}
\begin{equation}
 \zeta(q_{23},q_{{13}})=\frac{q_{{13}}^2+
 \lambda_1\lambda_2^{-1}(q_{23}^2
+\sigma\kappa^2)} {2\lambda_1q_{23}q_{{13}}} \label{21}
\end{equation}
in which $\kappa=\sqrt{2\mu_{(23)1}\varepsilon_a}$,
 $\sigma=\mu_{23}/\mu_{(23)1}= (1-\lambda_1\lambda_2)\lambda_2/\lambda_1$
and $Q_{L''}^{\eta}(x)$ is  the associate Legendre function of the
second kind \cite{Abr}. Note that $\zeta(q_{23},q_{{13}})>1$ . The
integration in Eq.(\ref{18}) has been performed by using the formula
(A.3) of Ref.\cite{Dol}. Now the integration over the angle
variables $\Omega_{{\bf {q}}_{23}}$ and $\Omega_{{\bf {q}}_{13}}$
can be performed by inserting the expressions (\ref{16}) and
(\ref{17}) in the r.h.s. of Eq.(\ref{3}). After that, the expression
(\ref{8}) for the three-body radial wave function with taking into
account (\ref{19}) can be reduced to
 the form
  $$ \Psi_{\nu}(r_{23},r_{{13}})=-\frac{m_3}{(2\pi)^4}
\frac{\sqrt{\hat{l}_{{13}}\hat{l}_{23}}}{r_{23}r_{13}}
(-1)^{l_{{13}}+l_{23}+L} \sum_{\hat{l^{\prime}}_{23}\hat{l^{\prime}}_{13}\hat{L^{\prime}}}
\hat{L^{\prime}}\sqrt{\hat{l^{\prime}}_{13}\hat{l^{\prime}}_{23}}
I_{l_{23}l_{13}L\hat{l^{\prime}}_{23}\hat{l^{\prime}}_{13}\hat{L^{\prime}}}(r_{23},r_{13})
$$
\begin{equation}
\times
\left( \begin{array}{ccc}
l_{23} & l_{23}' & L' \\
0 & 0 & 0
\end{array}\right )
\left( \begin{array}{ccc}
l_{{13}} & l_{{13}}' & L' \\
0 & 0 & 0
\end{array}\right )
\left\{\begin{array}{ccc}
l_{{13}} & l_{{13}}' & L' \\
l_{23}' & l_{23} & L
\end{array}\right \}\label{22}
\end{equation}
where  $\hat{j} =2j+1$  and
$$
I_{l_{23}l_{13}L\hat{l^{\prime}}_{23}\hat{l^{\prime}}_{13}\hat{L^{\prime}}}(r_{23},r_{{13}}) =
\int_{-\infty}^{\infty} dq_{23} e^{iq_{23}r_{23}}
f_{l_{23}}(q_{23}r_{23})
\int_{-\infty}^{\infty} dq_{{13}} e^{iq_{{13}}r_{{13}}}
f_{l_{{13}}}(q_{{13}}r_{{13}})
\left(\frac{q_{23}q_{{13}}}{4m_3\varepsilon_a}\right)^{\eta_a(q_{{13}})}
$$
\begin{equation}
\times
e^{i\pi\eta_a(q_{{13}})}\left(\zeta^2(q_{23},q_{{13}})-1\right)^{\eta_a(q_{{13}})/2}
Q_{L'}^{-\eta_a(q_{{13}})}[\zeta(q_{23},q_{{13}})]
W_{l_{23}'l_{{13}}'Ls_{12}S}(q_{23},q_{{13}}).\label{23}
\end{equation}
\vspace{0.5cm}

\hspace{0.2cm} The sought proper three-body asymptotic form for the radial wave function
$\Psi_{\nu}(r_{23},r_{{13}})$ at $r_{23}\to\infty$ (or $r_{{13}}\to\infty$) is
determined by the aforesaid three-body singularity  at the point
$\varepsilon=\varepsilon_a$. This singularity is associated with
  singularities of the function
$Q_{L'}^{-\eta_a(q_{{13}})}(\zeta(q_{23},q_{{13}}))$ entering  the integrand of Eq.(\ref{23}), and is   the power branch point \cite{Kr}\footnote{ Note that  character of this singularity is a logarithm
branch point one, when  values of the Coulomb parameter $\eta_a(q_{{13}})$ are integer numbers \cite{Be}.}. The latter singularity is defined from the equation
\begin{equation}
\zeta(q_{23},q_{{13}})=\pm 1.\label{24}
\end{equation}
The solution of Eq.(\ref{24}) results  in  the power branch point type of
singularities on
 the variable $q_{23}$ located at
  \begin{equation} \label{(2.21)}
 q_{23}^{(1,2)}=\pm\lambda_2q_{{13}}+i\sqrt{\sigma}\sqrt{q_{{13}}^2+\kappa^2}\label{25}
\end{equation}
and
\begin{equation}
\label{(2.22)}
q_{23}^{(3,4)}=\pm\lambda_2q_{{13}}-i\sqrt{\sigma}\sqrt{q_{{13}}^2+\kappa^2}\label{26}
\end{equation}
in the $q_{23}$ plane (see Fig.\ref{fig1}$a$).

To extract a contribution from these three-body singularities  to the radial wave
function $\Psi_{\nu}(r_{23},r_{{13}})$ given by   Eqs.(\ref{22}) and
(\ref{23}), first, a deformation of the contour  of integration into
the upper half of the $q_{23}$-plane is carried out as shown in Fig.\ref{fig1}$a$.
Then, in the integrand of the obtained integrals, the function
$e^{i\pi\eta_a(q_{{13}})}(\zeta^2-1)^{\eta_a(q_{{13}})/2}
Q_{L'}^{-\eta_a(q_{{13}})}(\zeta)$ is written as
$$
e^{i\pi\eta_a(q_{{13}})}(\zeta^2-1)^{\eta_a(q_{{13}})/2}Q_{L'}^{-\eta_a(q_{{13}})}
(\zeta)=\frac{\Gamma(L'-\eta_a(q_{{13}}))\
\Gamma(1-L'+\eta_a(q_{{13}}))}{2\Gamma(1+\eta_a(q_{{13}}))}
$$
\begin{equation}
\times
[(-1)^{L'}(\zeta-1)^{\eta_a(q_{{13}})}\,\,\,_2F_1(-L',L'+1;1+\eta_a(q_{{13}});\frac{1-\zeta}{2})-
\label{27}
\end{equation}
$$
-(\zeta+1)^{\eta_a(q_{{13}})}\,\,\,_2F_1(-L',L'+1;1+\eta_a(q_{{13}});
\frac{1+\zeta}{2})],
$$
where $_2F_1(a,b;c;x)$ is the hypergeometric function \cite{Abr} and
$\zeta\equiv \zeta(q_{23},q_{{13}})$. The relation (\ref{27}) can be
obtained from the formula 3.3.2(15) of Ref.\cite{Be}. As a result,
one can separate the parts of the integrals running along the cuts
$C^{(1)}$ and $C^{(2)}$ in the $q_{23}$ plane starting from
$q_{23}^{(1)}$ to $\infty$ and from $q_{23}^{(2)}$ to $-\infty$,
respectively, which correspond to the sought  asymptotics of the
three-body radial wave function $\Psi_{\nu}(r_{23},r_{{13}})$.  For
$r_{23}\to\infty$ one extracts  the contributions from the singular
points $q_{23}=q_{23}^{(1)}$ and $q_{23}=q_{23}^{(2)}$ in the
integrals over the contours $C^{(1)}$ and $C^{(2)}$, respectively.
After that, the expression (\ref{23}) can be reduced to the form
\begin{equation}
 I_{l_{23}l_{{13}}l_{23}'l_{{13}}'L'L}(r_{23},r_{{13}})
\approx \sum_{j=1}^2 (-1)^{j+1}
I_{l_{23}l_{{13}}l_{23}'l_{{13}}'L'L}^{(j)}(r_{23},r_{{13}}), \ \ \
\  r_{23}\rightarrow\infty\label{28}
\end{equation}
$$
I_{l_{23}l_{{13}}l_{23}'l_{{13}}'L'L}^{(j)}(r_{23},r_{{13}}) = \frac{\pi}{r_{23}}
\xi_{L'}^{(j)}\int_{-\infty}^{\infty} dq_{{13}}
e^{S^{(j)}(q_{{13}};r_{23},r_{{13}})}f_{l_{{13}}}(q_{{13}}r_{{13}})
$$
\begin{equation}
\times f_{l_{23}}(q_{23}^{(j)} r_{23})
\left(\frac{q_{13}^2+\kappa^2}{16\mu_{23}\mu_{(23)1}\varepsilon_a^2r_{23}^2}
\right)^{\eta_a(q_{{13}})/2}
W_{l_{23}'l_{{13}}'Ls_{12}S}(q_{23}^{(j)},q_{{13}}),\label{29}
\end{equation}
\begin{equation}
S^{(j)}(q_{{13}};r_{23},r_{{13}})=e^{iq_{{13}}r_{{13}}+iq_{23}^{(j)}r_{23}},\label{30}
\end{equation}
where $\xi_{L'}^{(1)}$=1 and $\xi_{L'}^{(2)}=(-1)^{L'}$. At the
limit $r_{23}\rightarrow\infty$, the integration over $q_{{13}}$ can
now be performed by using the saddle-point method \cite{Fed}. Before
applying this method, it should be noted that the function
$S(q_{23};r_{23},r_{{13}})$ in the exponent of the integrand of the
integral (\ref{29}) contains two parameters ($r_{23}$ and
$r_{{13}}$) and the  saddle-point method for such
 type of an integral has been developed in Ref.\cite{Fed}
(see for details Section 7.3 of Chapter 4 of Ref.\cite{Fed}).
According to Ref.\cite{Fed},  the saddle points in Eq.(\ref{29}) 
 determined from equation
\begin{equation}
\frac{dS^{(j)}(q_{{13}};r_{23},r_{{13}})}{dq_{{13}}}=0 \label{31}
\end{equation}
are  given by
\begin{equation}
q_{{13}}^{(j)} = i2\mu_{(23)1} \sqrt{\epsilon_a}\, \frac{\mid
r_{{13}}-(-1)^j
\lambda_2r_{23}\mid}{{R}^{(j)}(r_{23},r_{{13}})}, \label{32}
\end{equation}
where
$$
R^{(j)}(r_{23},r_{{13}})=\sqrt{2\mu_{23}r_{23}^2+2\mu_{(23)1}(r_{13}-(-1)^j
\lambda_2r_{23})^2}=
$$
\begin{equation}
= \sqrt{2\mu_{{13}}r_{{13}}^2+
2\mu_{({13})2}(r_{23}-(-1)^j\lambda_1r_{13})^2}\label{33}
\end{equation}
is a modified hyperradius \cite{Yar02}. It is noted that the
expressions $R^{(1)}$ and $R^{(2)}$ are respectively the maximum and
minimum values of the
 hyperradius $R$ \cite{Mer1} for fixed values of $r_{23}$ and $r_{{13}}$,
when the angle between the relative coordinates varies. As is seen from Eqs. (A11) and (A12) of Appendix A, they coincide with $R$
 when three
 particles are aligned: $R^{(1)}=R$   when
 particles 1 and 2
are on the opposite sides of a core 3 and $R^{(2)}=R$ when they are on
the same side. Inserting   expression (\ref{32}) into (\ref{25}), one obtains
 \begin{equation} q_{23}^{(j)} = i2\mu_{(13)2}
 \sqrt{\varepsilon_a}\frac{\mid r_{23}-(-1)^j\lambda_1
r_{13}\mid}{R^{(j)}(r_{23},r_{13})}.\label{34}
\end{equation}

As a result, from Eqs.(\ref{22}), (\ref{23}),  (\ref{28}) and (\ref{29}) the proper
three-body asymptotic form of the radial wave function
$\Psi_{\nu}(r_{23},r_{13})$ is derived  for $r_{23}\to\infty$ as
$$
\Psi_{\nu}(r_{23},r_{13})\simeq\Psi_{\nu}^{(as)}(r_{23},r_{13}) =
 \{C_{\nu}^{(1)}(r)f_{l_{23}}(q_{23}^{(1)}r_{23})
f_{l_{13}}(q_{13}^{(1)}r_{13})
[2\sqrt{\varepsilon_a} R^{(1)}(r_{23},r_{13})]^{-\eta_a(q_{13}^{(1)})}
$$
$$
\times
\exp[-\sqrt{\varepsilon_a} R^{(1)}(r_{23},r_{13})]/
[R^{(1)}(r_{23},r_{13})]^{3/2}
$$
$$ - C_{\nu}^{(2)}(r)
f_{l_{23}}(q_{23}^{(2)}r_{23}) f_{l_{{13}}}(q_{13}^{(2)}r_{13})
[2\sqrt{\varepsilon_a} R^{(2)}(r_{23},r_{13})]^{-\eta_a(q_{13}^{(2)})}
$$
\begin{equation}
\times \exp[-\sqrt{\epsilon_a} R^{(2)}(r_{23},r_{13})]/
[R^{(2)}(r_{23},r_{13})]^{3/2} \}/r_{23}r_{13},\label{35}
\end{equation}
where $\eta_a(q_{13}^{(j)})=iz_1z_3e^2\mu_{31}/q_{13}^{(j)}$. The
expression is also valid for $r_{13}\to\infty$ since the expressions
for  $q_{23}^{(j)}$, $q_{13}^{(j)}$ and $R^{(j)}$ do not change.
This means that the asymptotic formula is valid when
$r_{23}\to\infty$ and $r_{13}\to\infty$. However, similar to what it
has been  done in Ref.\cite{Yar02}, in the derivation of the saddle points
$q_{23}^{(j)}$ and $q_{13}^{(j)}$, one also implicitly assumes that
the ratio $r=r_{13}/r_{23}$ is larger than $\lambda_2$ and smaller
than $1/\lambda_1$.

In Eq.(\ref{35}), the asymptotic normalization functions
$C_{\nu}^{(j)}(r)$
 are related to the partial OSVF ,
$W_{l_{23}'l_{{13}}'Ls_{12}S}(q_{23}^{(j)},q_{13}^{(j)})$, by
$$
C_{\nu}^{(j)}(r)=N(-1)^{l_{13}+l_{23}+L}\sum_{l_{23}'l_{13}'L'}
\xi_{L'}^{(j)}\hat{L'}\sqrt{\hat l_{13}\hat l_{23}\hat l_{23}'\hat l_{13}'}
$$
\begin{equation}
\times
W_{l_{23}'l_{13}'Ls_{12}S}(q_{23}^{(j)},q_{13}^{(j)})
\left( \begin{array}{ccc}
l_{23} & l_{23}' & L' \\
0 & 0 & 0
\end{array}\right )
\left( \begin{array}{ccc}
l_{13} & l_{13}' & L' \\
0 & 0 & 0
\end{array}\right )
\left\{\begin{array}{ccc}
l_{13} & l_{13}' & L' \\
l_{23}' & l_{23} & L
\end{array}\right \},\label{36}
\end{equation}
where $N=(2\pi)^{-5/2} m_3(m_1m_2m_3/m)^{1/2}\varepsilon_a^{1/4}$.
It should be noted that for $\eta_a^{(j)}=0$ the expression (\ref{35}) coincides with that of Eq.(28)  obtained in Ref.\cite{Yar02} for three-body (123) bound system with pure short range nuclear two-particle interactions.
 
 As  is seen from Eqs. (\ref{35}) and (\ref{36}),  the asymptotics form for the three-body radial wave functions has the radial dependence quite  different from that for the total three-body wave function derived in Appendix A and given by  
 Eqs.   (A17)--(A19). Besides, the present asymptotic expression (\ref{35}) is not directly comparable to relation derived in Ref.\cite{Blokh05} for   the Jacobi variables  since the partial wave expansions are quite different.
But, when particle 3 is very heavy ($m_3\to\infty$, i.e.
$\lambda_1\to 0$ and $\lambda_2\to 0$), similar to what it is done
in \cite{Yar02}, it can easily be shown  that  the expression
(\ref{35}) becomes equivalent to the asymptotic formula (11) of
Ref.\cite{Blokh05}. When $m_3$ is not large, these two expressions
behave quite differently. The similar  situation occurs for the leading term of Eq.(\ref{35}) valid for $\mid q_{23}^{(j)}r_{23}\mid>>$1 and $\mid q_{13}^{(j)}r_{13}\mid>>$1 and the asymptotics form of  Eq.(A17) valid at $R\to\infty$  for the total three-body wave function. In this case,  both expressions have also the same radial dependence as $(2\sqrt{\varepsilon_a}R)^{-\eta_a}e^{-\sqrt{\varepsilon_a}R}/R^{5/2}$, where $\eta_a$ is given by Eqs. (A12), (A14) and (A19) of Appendix A at $\lambda_2$=0.   

As  is also seen  from Eq.(\ref{35}), the proper three-body  asymptotics
of the bound state radial wave function $\Psi_{\nu}(r_{23},r_{13})$ for $r_{23}\to \infty$
 and $r_{{13}}\to \infty$ contains unknown functions $C_{\nu}^{(j)}(r)$
 related to the partial OSVF $W_{\hat{l^{\prime}}_{23}\hat{l^{\prime}}_{13}Ls_{12}S}(q_{23}^{(j)},q_{13}^{(j)})$ by Eq.(\ref{36}). Consequently, as it was mentioned above,
 due to the fact that the partial OSVF is determined by the dynamics of strong
two-particles interaction \cite{Av}, knowledge of the factor
$C_{\nu}^{(j)}(r)$ allows one to get the valuable information both
on the three-body  structure of halo nuclei and on types of
two-particle (cluster-cluster, cluster-nucleon and nucleon-nucleon)
interactions.

\section{Analysis of the   three-body  wave
functions for  the second  excited   state  of the ${\rm{^6Li}}^*$ nucleus  }

\hspace{0.6cm} In this section,    the results of the comparative analysis of the  asymptotic expression    (\ref{35}) with  the approximate  radial three-body  wave functions  for the  halo  ($E^*$=3.562 MeV)
  state of ${\rm{^6Li}}$, which were derived by D. Baye \cite{Baye2007} within the framework of the $pn\alpha$ three-body approach based on the Lagrange-mesh technique [36--39] and given by Eq. (B2) in Appendix B,  are presented. These radial three-body  wave functions were obtained using  the central part of the Minnesota 
NN potential    and two kinds of the $\alpha N$ potential denoted by      the M1 and M2 models below (see, Appendix B also).  The leading asymptotic expression for the three-body
($pn\alpha$) radial wave functions for  ${\rm{^6Li}}$($E^*$=3.562 MeV) is determined by the expression
(\ref{35}),  since there are no two-particle bound subsystems for the
${\rm{^6Li}}$($E^*$=3.562 MeV) nucleus within the  three-body ($pn\alpha$) model.
 Note that, as it is mentioned above,  this second  excited  state of the  ${\rm{^6}}$Li nucleus  is  the isobar-analog   for
the ground state of the ${\rm{^6He}}$ halo nucleus. Therefore, similar to that   done in \cite{Yar02} for the  ground state of ${\rm{^6He}}$, it is of interest to find out  to what extent  the approximate radial wave functions   \cite{Baye2007}  have correct asymptotic behavior in the asymptotic region ($r_{23}\to\infty$ and $r_{13}\to\infty$).

\subsection{Results and discussion}
 
\hspace{0.6cm}   The TBANFs $C^{(j)}_{Ll}(r)$ ($j$=1
and 2) are determined based on the same  procedure  as used in Ref.\cite{Yar02}
for the ${\rm{^6He}}(nn\alpha)$ nucleus, where the index $\nu$ is substituted by   $ Ll$ since the quantum numbers summarized by index $\nu$ are fully determined by
the values $L$ and $l$. We consider value of  $r$ belonging
to   the interval 1.0 $\lesssim r\lesssim$ 3.0. This limit is connected with the fact that   other  values of $r$ correspond    to too large $r_{23}$ values (for $r<$ 1) for which the approximate wave function (B2) becomes inaccurate. For too small $r_{23}$ values (too large $r$ values)  the asymptotic expression (\ref{35}) is not valid. Therefore, the limit mentioned above   for $r$ (or $r_{23}$  and $r_{13}$) leads to the fact that  the   tail of the amplitude  of the three-body radial wave functions $\Psi_{Ll}(r_{23},r_{13})$ corresponds to values of $r_{23}\equiv r_{n\alpha}\gtrsim$ 5 fm and $r_{13}\equiv r_{p\alpha}\gtrsim$ 14 fm because of the small value of the binding energy $\varepsilon$ and  presence of the Coulomb $p\alpha$ interaction. 

After a number of tests, we have found that the convenient procedure
is to fix two coefficients $C^{(j)}_{Ll}(r)$ by fitting them to two
values of the partial wave function separated by a distance $\Delta
r_{23}$ for a fixed $r$. The optimal values of $\Delta r_{23}$ equal
to from  2.00 till 2.25 fm seem to be adequate. Similarly to Ref.
\cite{Yar02}, we have searched for a region where the obtained
values of the TBANF's $C^{(j)}_{Ll}(r)$ are stable within about
10\%, and have then tested the validity  of the obtained fit inside
that region.

As the examples of the search,  some values of $r$   are displayed in Table
\ref{table1} both for the M1 model and for the M2 model. The
obtained values are shown as a function of different choices for
fitting points ($r_{23}$,$r_{13}$) and ($r_{23}+\Delta
r_{23}$,$r_{13}+r\Delta r_{23}$) for the dominant partial waves
($L,l$)=(0,0),(0,1) and (0,2) for spin $S$=0, and ($L,l$)=(1,1) and
(1,2) for spin $S$=1. As   shown in Table \ref{table1}, different
choices of the fitting points can be found, which lead to similar values of the
coefficients $C^{(j)}_{Ll}(r)$.  The similar    situation occurs
  for other values of the ratio $r=r_{13}/r_{23}$.
Strikingly, the ratio $C^{(2)}_{Ll}/C^{(1)}_{Ll}$ is very stable. This indicates that only the overall normalization of the numerical wave function is  sensitive to the fitting points.   The width $\Delta$ ($\Delta=r_{23}^{(max)}-r_{23}^{(min)}$) of interval  for the variable $r_{23}$,   within which the approximate wave function (B2) has a correct asymptotic
behavior to within about 10\%, is presented in the sixth  and tenth columns of Table \ref{table1} for the M1  and M2 models, respectively.     Table \ref{table1}  shows that  the width of the interval becomes wider   with deceasing $r$  and more narrow    with increasing  ratio $r$ both for the M1 model  and the M2 model.  
  For example, for $L$=0, $l$=0 and  $r$=1.1, the expressions (\ref{35}) and (B2) agree within about 10\% for $r_{23}$   from about 14  to 24 fm for the M1 model and from about 14 to 27 fm for the M2 model. One notes that  the same behavior
is practically observed for $r_{13}=1.0r_{23}$. For     $r_{13}=2.5r_{23}$, the fit is excellent from about 6 to 11.5 fm for the M1 model and  from about 6 to 12.5 fm for the M2 model. One can observe  that  the width of
the interval  $\Delta$ becomes more narrow with increasing $r$ and
moves to smaller values of $r_{23}$. For the $L$=0 and $l$=1 partial wave  an accuracy of
about 10\% is obtained up to $r_{23}$ equal to from about 16  to about
23 fm for the M1 model and from about 16  to 24 fm  for the  M2 model at   $r_{13}=1.1r_{23}$.   The same accuracy is reached  from
about 6 to 11 fm for the M1 model   and from about 6 to 12 fm for  the M2 model  at $r_{13}=2.4r_{23}$. 
Here one observes that the width $\Delta$ deceases
 with increasing $r$ both for the M1 model and for the M2 model, but it moves to
smaller values   of $r_{23}$. For the small $L$=1 and  $l$=2 component    the quality of the agreement is reached from about 17 to 24  for the M1 model and about from 17  to 29 for the
 M2 model at $r_{13}=1.3r_{23}$. Similar situation is observed for other values of   $r$. To confirm  this and for the visual convenience, Fig. \ref{fig2} shows  the ratio of the approximate wave function (B2) to   the asymptotic form (\ref{35}) as a function of $r_{23}$   for different values of $r$, including the values of $r$ not  presented in Table \ref{table1}, only  for the M2 model.

It is seen here that, the width $\Delta$ depends both on values
of $r$ and on the form of the $\alpha N$ potential used. One also
observes in Table \ref{table1} that the region where the fit is
performed moves to larger distances when $l$ increases. This
reflects the fact that the centrifugal barrier  pushes the asymptotic
region to larger distances with increasing $l$. One notes that  the influence of the centrifugal barrier on  the asymptotic form (\ref{35}) arises only for $l>$ 0 and  is determined by the multiplicative factors $f_{l_{23}}(q_{23}^{(j)}r_{23}) f_{l_{{13}}}(q_{13}^{(j)}r_{23})$. These factors play an important role for the good agreement (within 10\%)  between the approximate  wave function (B2) and the asymptotic from (\ref{35}) in the asymptotic region,  similar to that  for the ${{^6He}}(nn\alpha)$ nucleus shown in  Refs. \cite{Yar02}.

 We have applied the fitting technique described above to the
determination of the TBANFs for  the   halo ($E^*$=3.562 MeV)  state
 of the ${\rm {^6Li}}$   nucleus
as a function of $r$ under the model conditions  described above. The recommended values of $C_{Ll}^{(j)}(r)$ for $j$=1 and 2 are presented
 in Figs. \ref{fig3} and \ref{fig4}    by the
solid and dashed curves   for the M1  and M2 models, respectively. 
Figs. \ref{fig3} and \ref{fig4} correspond
to   $L$=0 and 1, respectively. The curves   correspond to the averaged arithmetical means of
$C^{(j)}_{Ll}(r)$, which are obtained  from those for which the approximate wave function and the asymptotic form agree within about $\pm$10\% for the different
fitting points ($r_{23}$,$r_{13}$) and ($r_{23}+\Delta
r_{23}$,$r_{13}+r\Delta r_{23}$).  In all cases, one has $\mid C^{(1)}_{Ll}\mid>\mid C^{(2)}_{Ll}\mid$ similar to that found in \cite{Yar02} for the $^6$He(g.s.) nucleus.   One notes that the extracted values of the TBANFs $C^{(j)}_{Ll}(r)$ show a noticeable   sensitivity to the form of the  $\alpha N$ potential, although the binding energies calculated in the M1 and M2 models are fairly close to each other (see Appendix B).

\subsection{Comparison of the TBNFs for the  isobaric (${\rm{^6He}},{\rm{^6Li^*}}$) pair}

\hspace{0.6cm} It is now of interest to compare  the TBANF values  derived by us above for the leading components (($L\,l$)=(0,1) and (1,1)) and  the M2 model with those obtained   in Ref.\cite{Yar02} for the ground  
 state of $^6$He(g.s.). Note that, in Ref. \cite{Yar02} the TBANF values were   obtained only  for   the  nuclear 
 $\alpha N$ and $NN$ potentials used in  the  M2 model. For this purpose, we form the ratio $R_{L\,l}^{(j)}\equiv R_{L\,l}^{(j)}(r)=C^{(j)}_{Ll}({\rm{^6Li^*}})/C^{(j)}_{L\,l}({\rm{^6He}})$ as a function of $r$, where $C^{(j)}_{Ll}({\rm{^6Li^*}})$   denotes $C^{(j)}_{Ll}(r)$ for ${\rm{^6Li}}$($E^*$=3.562  MeV)  and $C^{(j)}_{L\,l}({\rm{^6He}})$ denotes $C^{(j)}_{L\,l}(r)$ for ${\rm{^6He}}$(g.s.). The results for the ratio are presented in Fig. \ref{fig5}. The  uncertainty for each  curve is about  $\pm$14\%, which is the average square error for the ratio $R_{L\,l}^{(j)}$, which involves  the uncertainties of the $C^{(j)}_{Ll}({\rm{^6Li^*}})$   and $C^{(j)}_{L\,l}({\rm{^6He(g.s.)}})$.  As is seen from figures, the interval of changing the variable $r$ can be divided in three parts (denoted by $\Omega_f,\,\,f$=1, 2 and 3).
 In $\Omega_1$, a value of $r$ changes from 1.0 to about 1.1 for which   the values of $R_{L\,l}^{(j)}$ are noticeably larger than   1.  These values of $r$  correspond to  rather  large values for the ($r_{23},r_{13}$) pair, which are   the   fitting points    providing the stable values of the  $C^{(j)}_{Ll}({\rm{^6Li^*}})$.   For example,  $R_{0\,1}^{(1)}$=2.08$\pm$0.29 and $R_{0\,1}^{(2)}$=1.98$\pm$0.28  as well as  $R_{1\,1}^{(1)}$=1.73$\pm$0.24 and  $R_{1\,1}^{(2)}$=1.62$\pm$0.23  at   ($r_{23},r_{13}$)=(19.0 fm,19.0 fm) for $r$=1.0. In $\Omega_2$, a value of $r$ changes within the interval  1.3$\lesssim r\lesssim$1.7 for which   the values of $R_{L\,l}^{(j)}$ are quite close to  1. In $\Omega_2$, values of  $r_{23}$ and $r_{13}$ decrease for the fitting  points      providing the stable values of the TBANFs $C^{(j)}_{Ll}({\rm{^6Li^*}})$.   For example,  $R_{0\,1}^{(1)}$=1.11$\pm$0.16 and $R_{0\,1}^{(2)}$=0.99$\pm$0.14  as well as  $R_{1\,1}^{(1)}$=1.18$\pm$0.17 and  $R_{1\,1}^{(2)}$=1.00$\pm$0.14 at   ($r_{23},r_{13}$)=(13.50 fm,17.55 fm) for $r$=1.3. At last, in $\Omega_3$,   the values of $R_{L\,l}^{(j)}$ are noticeably less than  1 for $r\gtrsim$1.9. 
 For  these values   $r$ and     the fitting points  for the ($r_{23},r_{13}$) pair,        the values of  $r_{23}$ decrease and values of $r_{13}$ increase. For example,  $R_{0\,1}^{(1)}$=0.81$\pm$0.11 and $R_{0\,1}^{(2)}$=0.69$\pm$0.10  as well as  $R_{1\,1}^{(1)}$=0.75$\pm$0.10 and  $R_{1\,1}^{(2)}$=0.63$\pm$0.10 at   ($r_{23},r_{13}$)=(9.00 fm,18.9 fm) for $r$=2.1.

 As  is seen from here, the mirror symmetry for the TBANFs derived for  the isobar-analog states  of the  isobaric  (${\rm{^6He}}{\rm{^6Li}}^*$) pair breaks up significantly in $\Omega_1$ and $\Omega_3$. Apparently, this is  connected with the fact that  an  influence  of the Coulomb $p\alpha$ interaction upon  the derived  TBANFs $C^{(j)}_{Ll}({\rm{^6Li^*}})$ becomes noticeable for the values of $r_{13}$ changing in $\Omega_1$ and $\Omega_3$.     Nevertheless, in $\Omega_2$, the mirror symmetry for the TBANF values
derived for $C_{Ll}^{(j)}(^6Li^*)$ and $C_{Ll}^{(j)}(^6He)$ occurs within
their uncertainties.  One of the main reason is  the fact that the values of $r_{13}$(=$r_{p\alpha}$) for the fitting points in $\Omega_2$ are less than those in  $\Omega_1$ and $\Omega_3$ and, consequently, an  influence  of the Coulomb $p\alpha$ interaction decreases in  $\Omega_2$ with respect to that of the Coulomb $p\alpha$ interaction  in $\Omega_1$ and $\Omega_3$.

\section{Conclusion}

\hspace{0.8cm}In this work, we have determined the asymptotic  forms
of the radial and total three- body (123) wave functions of halo nuclei
with two charged particles (1 and a core 3) in core-nucleon coordinates. Although,
such expressions are already known in hyperspherical (or Jacobi)
coordinates \cite{Blokh05,Mukh}, we think that the present formulae are interesting
because they allow a simpler physical visualization of the
asymptotic behavior as a function of the two core-nucleon relative  distances $r_{23}$ and
$r_{13}$. It is demonstrated that when   the core  is very heavy the
derived asymptotic forms become equivalent to the corresponding
asymptotic formulae \cite{Blokh05,Mukh} for the Jacobi coordinates.
 
The asymptotic form obtained   for the radial wave
functions  has been compared  in different parts of  asymptotic regions with the   approximate wave functions derived by D. Baye  \cite{Baye2007}  within the Lagrange-mesh approach for the     halo  ($E^*$=3.562 MeV) state of $^6$Li($pn\alpha$) using two kinds of the nuclear $\alpha N$ potential. The intervals of the $r_{23}$ and $r_{13}$ variables within which  the
approximate wave functions have a correct asymptotic behavior within
about 10\% have been determined.

One has used the region where the agreement between a radial wave
function and the asymptotic expressions is excellent to deduce
values  for the  TBANFs, $C^{(1)}_{L\,l}(r)$ and $C^{(2)}_{L\,l}(r)$,
 for the ${\rm{^6Li^*}}$($E^*$=3.562 MeV)   nucleus depending on  the ratio $r=r_{13}/r_{23}$. The shape of the asymptotic behavior is determined by the ratio
$C^{(2)}_{L\,l}(r)/C^{(1)}_{L\,l}(r)$ and the overall normalization  
is sensitive to the fitting points. It is demonstrated that the values of the
 TBANFs are   sensitive to the form of the   $\alpha N$ potential used. Besides,  it is revealed   the region of changing   the ($r_{n\alpha},r_{p\alpha}$) pair where the mirror symmetry occurs for the TBANFs derived for the isobaric (${\rm{^6He}}$(g.s.),${\rm{^6Li^*}}$($E^*$=3.562 MeV)) pair.

The deduced three-body   asymptotic functions are in principle
observable quantities, for example, from an analysis of the
experimental differential cross sections for the exchange $\alpha
({\rm{^6Li}},\alpha){\rm{^6Li}}$(3.562 MeV)  and transfer  $\alpha
({\rm{^3He}},p){\rm{^6Li}}$(3.562 MeV) reactions.
Therefore, it would be interesting to compare  present results
with experimental ones. It would make it possible    to choose the
form of the $\alpha N$ potential by comparing the phenomenological
values of $C^{(j)}_{Ll}(r)$ ($j$=1 and 2) with these values obtained
in the present work. They would allow one to get additional
information about the form of the $\alpha N$ potential  and to
verify an accuracy of the approximate wave function (B2) as a
source of reliable information on the $C^{(j)}_{Ll}(r)$ asymptotic
normalization functions. In this connection, it would be highly
encouraged to do such   an experiment.

\begin{center}
ACKNOWLEDGEMENTS
\end{center}

The author is deeply grateful to D. Baye and L.  D. Blokhintsev for   reading a manuscript, discussions and general encouragement. The author thanks Y. Suzuki for interest and  D. Baye for providing the   numerical  data calculated for the   approximate three-body radial wave functions for $^6Li$($E^*$=3.562 MeV). The work has been  supported   by  The Academy of Sciences of   the Republic of Uzbekistan under grant No.  F2-FA-F177.
 
\newpage

\vspace{0.3cm} \begin{center}{\bf APPENDIX A: ASYMPTOTIC BEHAVIOR OF
THE FULL THREE-BODY WAVE FUNCTION}\end{center} 
\vspace{0.5cm}

\hspace{0.2cm} Here, it is interesting to derive a proper three-body
asymptotics  of the total three-body wave function $\Psi({\mathbf
r}_{23},{\mathbf r}_{13})$ at $r_{23}\to\infty$ (or
$r_{13}\to\infty$) for the  bound three-body ($a$=(123)) state  directly using   Fourier representation for the total three-body wave function.
 But, first one notes that in   Refs.\cite{Mer1,Mukh} an
asymptotic behavior of the total three-body wave function
$\Psi({\mathbf R})$  for $R\to\infty$ has been investigated by using
the Fourier representation  written through the six-dimensional
vectors  ${\mathbf R}=\{{\mathbf x},{\mathbf y}\}$[=$\{R,\hat {{\mathbf
R}}\}$] and ${\mathbf P}=\{{\mathbf q},{\mathbf p}\}$[=$\{P,\hat
{{\mathbf P}}\}$]. Here $x=\sqrt{2\mu_{12}
r_{12}} $ and $y=\sqrt{2\mu_{(12)3}
r_{(12)3}}\,\, $ (${\mathbf q}$ and ${\mathbf p}$) are the modified
coordinate Jacobi variables (conjugate
 momentums to them) \cite{Mer1}; $ \hat {{\mathbf R}}=\{\hat {{\mathbf x}},\hat {{\mathbf
y}},\varphi \}$; $R=\sqrt{x^2+y^2}$ and $\varphi=\arctan(y/x)$ are the
hyperradius and hyperangle in the configuration space, respectively; ${\mathbf
r}_{(12)3}$ is the radius vector connecting the centers of masses of the
(12)-pair and the particle 3; $P=\sqrt{q^2+p^2}$ and $\hat {{\mathbf
 b}}={\mathbf b}/b$ is a unit radius vector. In particular,
it was shown in work of Ref.\cite{Mukh} that the derived asymptotic
form $\Psi({\mathbf R})$  at $R\to\infty$ contains a factor of
OSVF for the virtual decay (11), which depends on the variables
$\varphi$, $\hat {{\mathbf x}}$ (or $\hat {{\mathbf r}}_{12}$)
   and  $\hat {{\mathbf y}}$ (or $\hat {{\mathbf r}}_{(12)3}$). Besides, in Ref.\cite{Mukh} the important relation between the total  OSVF  and the total three-body asymptotic normalization coefficient $C(\hat {\bf R})$ has also been found.

The  Fourier  transformation   for the total wave function
 is written as
$$
\Psi({\mathbf r}_{23},{\mathbf r}_{13})=\int \frac{d{\mathbf
q}_{23}}{(2\pi)^3}\frac{d{\mathbf q}_{13}}{(2\pi)^3}\,\,
e^{i({\mathbf r}_{23}{\mathbf q}_{23}+{\mathbf r}_{13}{\mathbf
q}_{13})} \Psi({\mathbf q}_{23},{\mathbf q}_{13}) \eqno(A1)
$$
$$
=-\int\frac{d{\mathbf q}_{23}}{(2\pi)^3}\,\,e^{i({\mathbf
r}_{23}-\lambda_1{\mathbf r}_{13}){\mathbf q}_{23}}I({\mathbf
q}_{23}; {\mathbf r}_{13}). \eqno(A2)
$$
Herein  the expressions  (\ref{12}), (\ref{14}) and (\ref{15}) as well as the substitution ${\mathbf q}_{13}'={\mathbf q}_{13}+\lambda_1{\mathbf q}_{23}$  in Eq.(A1)  are used,  where
 $$ I({\mathbf q}_{23};{\mathbf r}_{13})= \frac{1}{4\varepsilon_a}
\int\frac{d{\mathbf q}_{{13}}'}{(2\pi)^3}\,\,e^{i{\mathbf r}_{13}
{\mathbf q}_{13}'} \Gamma(1-\eta_a({\mathbf q}_{13}^\prime,{\mathbf
q}_{23})) $$ $$
 \times W_a(-{\mathbf
q}_{23},-{\mathbf q}_{13}'+\lambda_1{\mathbf q}_{23}) \left(\frac{q_{13}^{\prime
 2}/2\mu_{31}+ q_{23}^2/2\mu_{(31)2}+\varepsilon_a}
{4\varepsilon_a}\right)^{\eta_a(\bf {q}_{13}^\prime,\bf {q}_{23})-1}
 \eqno(A3)
$$
and
$\eta_a(q_{{13}})\equiv\eta_a({\mathbf q}_{13}^\prime,{\mathbf q}_{23})= iz_3 z_1
e^2\mu_{31}/\mid {\mathbf q}_{13}'-\lambda_1{\mathbf q}_{23}\mid$.

\begin{figure}
\begin{center}
\epsfxsize=15.cm \centerline{\epsfbox{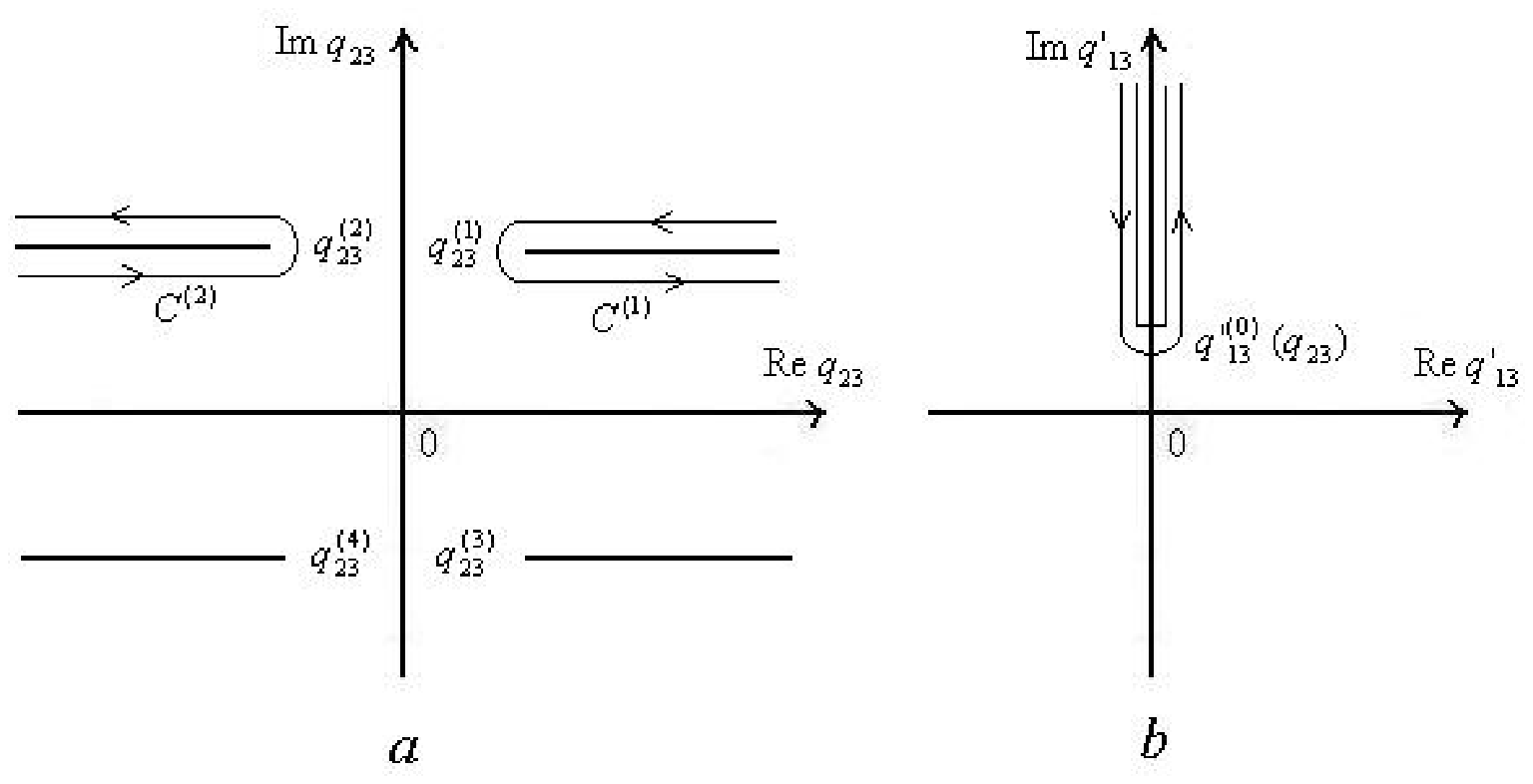}}
\caption{\label{fig1}Contours in the $q_{23}$ and $q^{\prime}_{13}$ complex planes.}
\end{center}
\end{figure}

\begin{figure}
\begin{center}
\epsfxsize=15.cm \centerline{\epsfbox{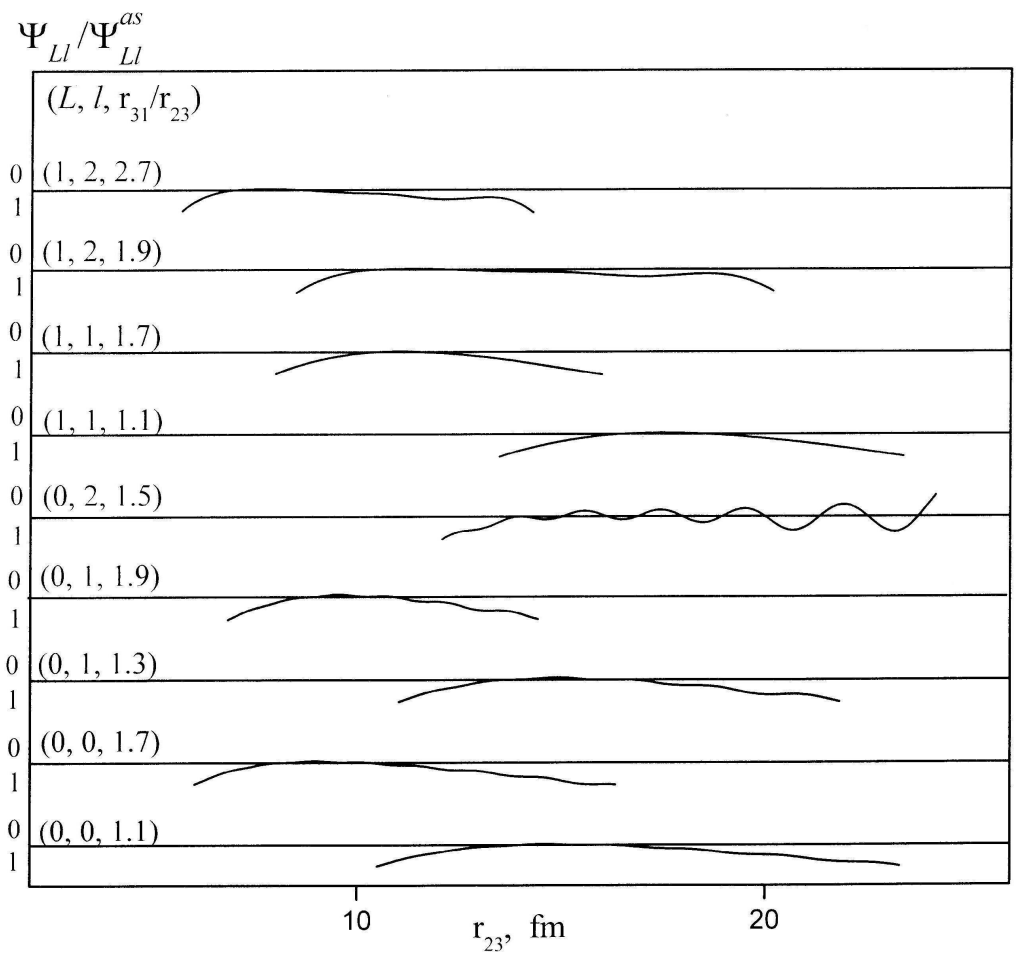}}
\caption{\label{fig2} Ratio of the  numerical partial wave functions (\ref{38}) to the asymptotic expression (\ref{35})  as a function of $r_{23}$ for fixed values of the coordinate ratio $r=r_{13}/r_{23}$. Each case is calculated for the ($r_{23},r_{13}$ ) pair of fitting points which provides the corresponding  values   $C^{(j)}_{Ll}(r)$ obtained by means of the fitting procedure mentioned above.  }
\end{center}
\end{figure}

\begin{figure}
\begin{center}
\epsfxsize=15.cm \centerline{\epsfbox{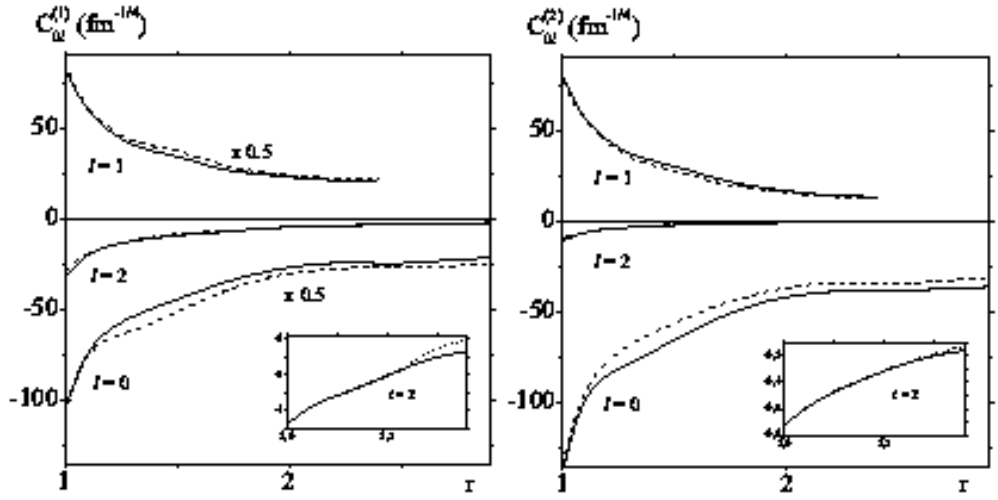}}
\caption{\label{fig3} The recommended values of the function $C_{Ll}^{(j)}(r)$ for $L$=0 and $l$=0, 1 and 2 ($j$=1 and 2).  The solid and dashed lines correspond to the M1 and M2 models, respectively. The uncertainty  for each the curve is within up to $\pm$10\%. }
\end{center}
\end{figure}

\begin{figure}
\begin{center}
\epsfxsize=15.cm \centerline{\epsfbox{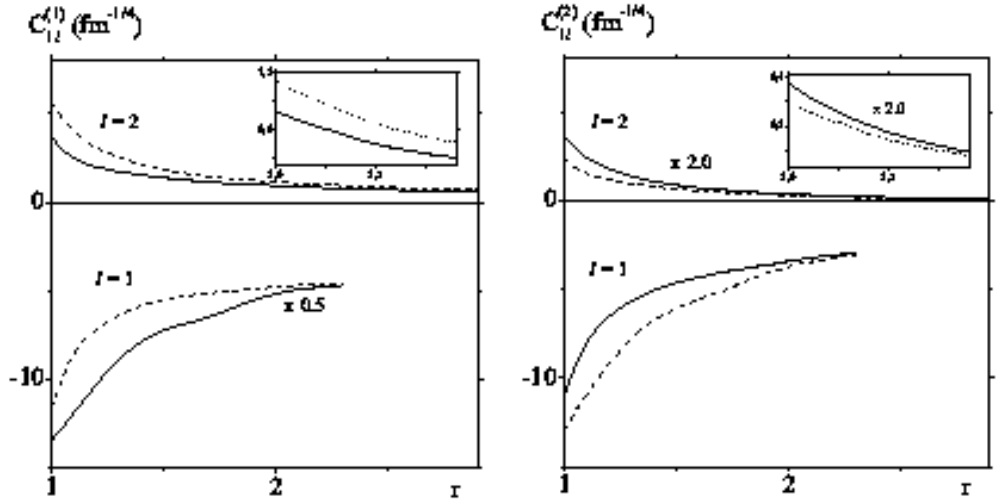}}
\caption{\label{fig4}Same as Fig.2 for $L$=$S$=1 and $l$=1 and 2.}
\end{center}
\end{figure}

\begin{figure}
\begin{center}
\epsfxsize=15.cm \centerline{\epsfbox{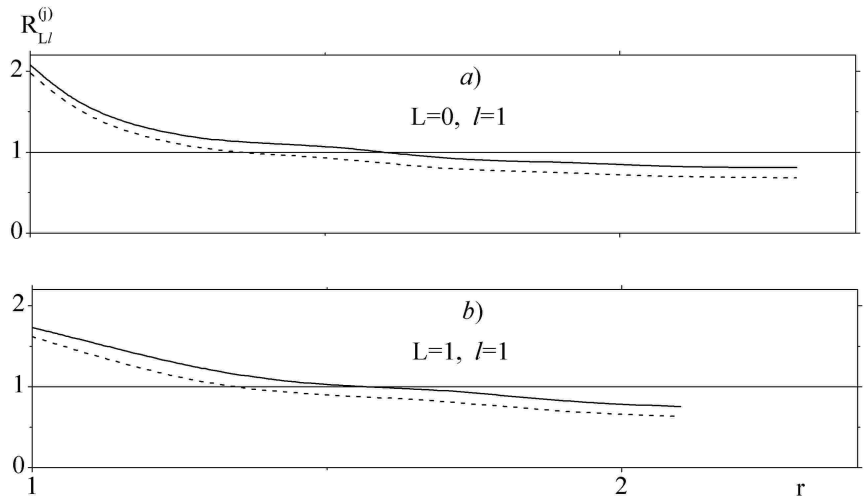}}
\caption{\label{fig5}The ratio $R_{L\,l}^{(j)}$ of the TBANFs derived for  the ${\rm{{^6Li}}}$(3.562 MeV,
$J^\pi$=0$^+$, $T$=1) nucleus to those for  ${\rm{{^6He}}}$(g.s., $J^\pi$=0$^+$, $T$=1) nucleus  as a function of $r$  for the M2 model. The solid and dashed lines correspond to $j$=1 and 2, respectively. The data for  ${\rm{{^6He}}}$(g.s., $J^\pi$=0$^+$, $T$=1) are taken from \cite{Yar02}. }
\end{center}
\end{figure}

We consider the expression (A3) at $r_{13}\to\infty$ and perform
integration over ${\mathbf q}_{13}'$ in it. Similar to
Refs.\cite{Mer1,Mer2}, the leading term of the integrals over
${\mathbf q}_{13}'$ at $r_{13}\to\infty$ is generated by the
contribution from the three-body branch point singularity located at
$\varepsilon=-\varepsilon_a$. Similar to what  has been done in
Ref.\cite{Mer2} (see for details, e.g., Section 6 of the Chapter 4
there) an estimation of the leading term of the integrals over
${\mathbf q}_{13}'$ at $r_{13}\to\infty$ can be done by using the
multidimensional method of stationary phase \cite{Fed}.  The
stationary phase points are given by $\hat {{\mathbf
q}}_{13}^{\prime}=\pm\hat {{\mathbf r}}_{13}$. When the expression
(A3) can be reduced  to the form
$$
I({\mathbf q}_{23}; {\mathbf r}_{13})\simeq I^{(as)}({\mathbf q}_{23};{\mathbf
r}_{13}) =\frac{1}{16i\pi^2\varepsilon_a r_{13}}
\int_{-\infty}^{\infty}dq_{13}'q_{13}'e^{iq_{13}'r_{13}}
\Gamma(1-\eta_a(\hat {{\mathbf r}}_{13} q_{13}^\prime,{\mathbf q}_{23}))
$$
$$
\times
W_a(-{\mathbf q}_{23},-\hat{{\mathbf r}}_{13} q_{13}^\prime+
\lambda_1{\mathbf q}_{23})\left
(\frac{q_{13}^{\prime 2}/2\mu_{31}+ q_{23}^2/2\mu_{(31)2}+\varepsilon_a}
{4\varepsilon_a}\right)^{\eta_a(\hat{{\mathbf r}}_{13}q_{13}^\prime,
{\mathbf q}_{23})-1}
\eqno(A4)
$$
 for $r_{13}\to\infty$,  where $\eta_a(\hat {{\mathbf r}}_{13}q_{13}^\prime,
{\mathbf
q}_{23})= iz_3z_1 e^2\mu_{31}/\mid\hat{{\mathbf r}}_{13} q_{13}'
-\lambda_1{\mathbf q}_{23}\mid$.
Then, one extracts contribution from the three-body branch point  singularity
 located at $q'_{13}=iq'^{(o)}_{13}(q_{23}) 
\equiv i\sqrt{2\mu_{31}(q_{23}^2/2\mu_{(31)2}+\varepsilon_a)}$
of the integrand in  (A4) by means of a deformation of the contour
 of integration into the upper half of the $q_{13}^\prime$ plane as plotted in Fig.{\ref{fig1}$b$.
 As a result, one can separate the part  of integral running
along the cut lying on the imaginary axis in the $q_{13}^\prime$ plane starting from
$q'^{(o)}_{13}(q_{23})$ to $\infty$ that corresponds to the leading term of the
expression (A4) at $r_{13}\to\infty$.
After that, expression (A4) can be reduced to the form
$$
 I({\mathbf q}_{23};{\mathbf r}_{13})\simeq
I^{(as)}({\mathbf q}_{23};{\mathbf r}_{13}) \simeq
\frac{1}{2\pi}\frac{\mu_{13}}{r_{13}}e^{ir_{13}q'^{(+)}(q_{23})}
$$
$$
\times
\left[\frac{q'^{(+)}(q_{23})}{4\mu_{31}
\varepsilon_ar_{13}}\right]^{\eta_a(\hat{{\mathbf r}}_{13}q'^{(+)}(q_{23}),{\mathbf
q}_{23})} W(-{\mathbf q}_{23},-\hat{{\mathbf r}}_{13} q'^{(+)}(q_{23})
+\lambda_1{\mathbf q}_{23}),\,\,\,\,\, r_{13}\to\infty.
\eqno(A5)
$$

By inserting the expression (A5) into the r.h.s. of Eq.(A2), the
latter can be rewritten to the form
$$
\Psi({\mathbf r}_{23},{\mathbf r}_{13})\simeq \Psi^{(as)}({\mathbf r}_{23},
{\mathbf r}_{13})=
\int d{\mathbf q}_{23}e^{r_{13}S({\mathbf q}_{23};{\mathbf r}_{23}, {\mathbf r}_{13})}
h({\mathbf q}_{23};{\mathbf r}_{23}, {\mathbf r}_{13})
\eqno(A6)
$$
for $r_{13}\to\infty$. Herein
$$
h({\mathbf q}_{23};{\mathbf r}_{23}, {\mathbf r}_{13})=-\frac{1}{(2\pi)^4}\frac{\mu_{31}}
{r_{13}}
$$
$$
\times\left(\frac{q_{23}^2/2\mu_{(31)2}+\varepsilon_a}{8\mu_{31}\varepsilon_a^2
r_{13}^2}\right)^{\eta_a(\hat {\bf {r}}_{13}q'^{(+)}(q_{23}),\bf
{q}_{23})/2} W(-{\mathbf q}_{23},-\hat {{\mathbf r}}_{13}
q'^{(+)}(q_{23}) +\lambda_1{\mathbf q}_{23}), \eqno(A7)
$$
$$
S({\mathbf q}_{23};{\mathbf r}_{23}, {\mathbf r}_{13})=i({\mathbf r}_{23}-
\lambda_1{\mathbf r}_{13})
{\mathbf q}_{23} /r_{13}- q^{\prime (+)}(q_{23}).
\eqno(A8)
$$

At $r_{13}\to\infty$, the integration over $\bf {q}_{23}$ can be performed
 by using the multidimensional saddle-point method \cite{Fed}. The saddle points
are determined from   equations
$$
 grad_{{\mathbf q}_{23}} S({\mathbf q}_{23};{\mathbf r}_{23}, {\mathbf r}_{13})=0.
\eqno(A9)
$$
The saddle points are given by
$$
{\mathbf q}_{23}={\mathbf q}_{23}^{(o)}\equiv i2\mu_{(31)2}\sqrt{\varepsilon_a}
\frac{{\mathbf r}_{23}-\lambda_1{\mathbf r}_{13}}{R({\mathbf r}_{23}
,{\mathbf r}_{13})},
\eqno(A10)
$$
where
$$
R({\mathbf r}_{23},{\mathbf r}_{13})=\sqrt{2\mu_{31}r_{13}^2+
2\mu_{(31)2}({\mathbf r}_{23}-\lambda_1{\mathbf r}_{13})^2}
\eqno(A11)
$$
$$
\equiv \sqrt{2\mu_{23}r_{23}^2+2\mu_{(23)1}({\mathbf r}_{13}-\lambda_2{\mathbf
r}_{23})^2}
\eqno(A12)
$$
is the  hyperradius $R$. Whereas
$$
 {\mathbf q}_{13}={\mathbf
q}_{13}^{(o)}\equiv \hat {{\mathbf r}}_{13}q'^{(o)}_{13}(q_{23}^{(o)})-
\lambda_1{\mathbf q}_{23}^{(o)}
\eqno(A13)
$$
 at $\varepsilon=\varepsilon_a$,  from the relations (A10) and (A13), one
obtains
$$
{\mathbf q}_{13}^{(o)}=i2\mu_{(23)1}\sqrt{\varepsilon_a}
\frac{{\mathbf r}_{13}-\lambda_2{\mathbf r}_{23}}{R({\mathbf
r}_{23},{\mathbf r}_{13})}. \eqno(A14)
$$

 An asymptotic expression sought by us is generated by contributions from the
saddle points  ${\mathbf q}_{23}={\mathbf q}_{23}^{(o)}$. To this
end, we can employ the following standard techniques in the r.h.s.
of (A6). Firstly, the part $C$ of the integration region  
containing the vicinity of the saddle points ${\mathbf
q}_{23}={\mathbf q}_{23}^{(o)}$ is separated, i.e. $\{
q_{23;1}^{(o)},q_{23;2}^{(o)},q_{23;3}^{(o)}\} \in C$, where
$q_{23;j}^{(o)}$ ($j$=1,2 and 3) is a corresponding coordinate of the
vector ${\mathbf q}_{23}^{(o)}$. Then, the leading term of the
r.h.s. of (A6) at $r_{13}\to\infty$ can  be found by using the
following  formula (1.10$^{\prime}$) of Section 3 of the Chapter 4
of Ref.\cite{Fed}:
$$
\int d{\mathbf q}_{23}e^{r_{13}S({\mathbf
q}_{23};{\mathbf r}_{23}, {\mathbf r}_{13})} h({\mathbf q}_{23};
{\mathbf r}_{23}, {\mathbf
r}_{13})\simeq \int_C d{\mathbf q}_{23}e^{r_{13}S({\mathbf q}_{23};
{\mathbf r}_{23}, {\mathbf r}_{13})}
h({\mathbf q}_{23};{\mathbf r}_{23}, {\mathbf r}_{13})
 \eqno(A15)
$$
$$
\simeq
\left (\frac{2\pi}{r_{13}}\right )^{3/2} {e}^{r_{13}S({\mathbf
q}_{23}^{(o)};{\mathbf r}_{23}, {\mathbf r}_{13})} \left [ \det\left
\Vert-\frac{\partial^2}{\partial q_{23;i} q_{23;j}}
S\left ({\mathbf q}_{23};{\mathbf
r}_{23}, {\mathbf r}_{13}\right )\right \Vert _{{\mathbf q}_{23}=
{\mathbf q}_{23}^{(o)}}
\right]^{-1/2}
$$
$$
\times h({\mathbf q}_{23}^{(o)};{\mathbf r}_{23},{\mathbf r}_{13})
\eqno(A16)
$$
for $r_{13}\to\infty$, where $q_{23;j}$
($j$=1, 2, and 3) is a corresponding coordinate of the vector ${\mathbf q}_{23}$.

As a result, from Eqs.(A6)--(A8), (A15) and (A16) the proper
three-body asymptotics of the total wave function $\Psi({\mathbf
r}_{23},{\mathbf r}_{13})$ $r_{13}\to\infty$ can be obtained  as
$$
\Psi_a({\mathbf r}_{23},{\mathbf r}_{13})\simeq
\Psi_a^{(as)}({\mathbf r}_{23},{\mathbf r}_{13})=
C_a(r,\hat{{\mathbf r}}_{23},\hat{{\mathbf r}}_{13})(2\sqrt{\varepsilon_a}
R)^{-\eta_a(r,\hat{{\mathbf r}}_{23},\hat{{\mathbf r}}_{13})}
$$
$$
\times
e^{-\sqrt{\varepsilon_a}R}/R^{5/2}.
\eqno(A17)
$$
Herein
$$
 C_a(r,\hat{{\mathbf r}}_{23},\hat{{\mathbf r}}_{13})=
 -\frac{(m\sqrt{\varepsilon_a})^{3/2}} { \sqrt2 \pi^{5/2}}
W(-{\mathbf q}_{23}^{(o)},-{\mathbf q}_{13}^{(o)})
\eqno(A18)
$$
is the three-body asymptotic normalization factor,
$R\equiv R({\mathbf r}_{23},{\mathbf r}_{13})$ and
$$
\eta_a(r,\hat{{\mathbf r}}_{23},\hat {{\mathbf r}}_{13})=iz_3z_1e^2
\mu_{{13}}/q_{{13}}^{(o)}.
\eqno(A19)
$$
Recall that the function $W(-{\mathbf q}_{23}^{(o)},-{\mathbf q}_{13}^{(o)})$
 is the total OSVF for the virtual decay (11) because when ${\mathbf
q}_{23}={\mathbf q}_{23}^{(o)}$ and ${\mathbf q}_{{13}}={\mathbf
q}_{{13}}^{(o)}$, the particles 1, 2 and 3 are on the mass shell.

 It should be noted that the  asymptotic formula  is also valid for
$r_{23}\to\infty$. This point can be easily shown if  permutation of
the order of integration over ${\mathbf q}_{13}$ and ${\mathbf
q}_{23}$ is done in (A1). As a result, one obtains the same asymptotic formula (A17). This means
that asymptotic formula (A17) is valid when both $r_{23}$ and
$r_{13}$ (or ${R}$) tend to infinity.

The expression (A17) coincides with the asymptotic expression
obtained in Ref.\cite{Mukh} for $\Psi({\mathbf R})$ at $R\to\infty$
if in the latter a value of the charge ($z_b\equiv z_2$) for the
particle 2 is put  to zero, the three-body asymptotic normalization factor
$C(\hat {{\bf R}})$ is substituted by $C(r,\hat {{\bf r}}_{23},\hat
{{\bf r}}_{13})$ and the hyperradius $R$ is determined by either
Eq.(A11) or Eq.(A12). It follows from here that the limit
$R\to\infty$ for the asymptotic expression obtained in
Ref.\cite{Mukh} indeed means that both $r_{23}$ and $r_{13}$ tend
simultaneously to infinity.  
\newpage

\vspace{0.3cm} 
\begin{center}{\bf APPENDIX B: APPROXIMATE   
  THREE-BODY WAVE FUNCTIONS}\end{center} 
\vspace{0.5cm}

Below,  we briefly  present  the idea and the essential formulas of   
the aforementioned Lagrange-mesh technique utilized by D. Baye  \cite{Baye2007} for the  halo  ($E^*$=3.562 MeV}) state of ${\rm{^6Li}}$.

 According to Ref.\cite{Suz91}, the Hamiltonian
describing the relative motion of the nucleons with respect to the
core ($\alpha$-particle) reads
$$
H=\frac{\hat {\mathbf {q}}_{p\alpha}^2}{2\mu_{p\alpha}}+\frac{{\hat
{\mathbf {q}}}_{n\alpha}^2}{2\mu_{n\alpha}}+\frac{{\hat {\mathbf
{q}}}_{p\alpha}{\hat {\mathbf {q}}}_{n\alpha}
}{5\mu_{p\alpha}}+V_{p\alpha}+V_{n\alpha}^N+V_{np}^N,
 \eqno(B1)
$$
where   $V_{ij}=V_{ij}^N+V_{ij}^C$ and $V_{ij}^N(V_{ij}^C)$ is the
nuclear(Coulomb) potential between the centers of mass of particles
$i$  and $j$. The eigenvalue of $H$ provides a binding energy of the
${\rm{^6Li}}$($E^*$=3.562) state in the
($p+n+\alpha$)-channel. 
According  to \cite{Baye2007},    the central part of the Minnesota 
NN potential with an exchange parameter $u$=1 \cite{Tang78}   and two kinds of the $\alpha N$ potential  were employed. The latters  are taken from Ref.\cite{Voron95} (the M1 model), 
which takes into account the exchange Majorana component both in the
central and the spin-orbit term, and from Ref.\cite{Kan79} (the M2
model) with the the central and the spin-orbit terms.  In (B1),
they are both expressed as sums of Gaussians, in which,  similar to 
Ref.\cite{Kras75},  the pseudopotential technique is also applied in \cite{Baye2007} to  eliminate   the forbidden $s$ state of the $\alpha N$ interaction, which stimulates the Pauli antisymmerization principle between the nucleon and the core.

Within the framework of the Lagrange-mesh technique [36--39],
  a partial wave of the $pn\alpha$ wave
function for the second excited
 state of the ${\rm{^6Li}} $ nucleus  is presented as
$$
\Psi_{lLS}(r_{23},r_{13})=(r_{23}r_{13})^{-1}\sum_{i_1,i_2=1}^N
c_{i_1i_2}^{lLS}F_{i_1i_2 }(r_{23},r_{13}). \eqno(B2)
$$
In this expression, the $F_{i_1i_2 }$ are Lagrange basis functions
and the $c_{i_1i_2}^{lLS}$ are variational coefficients. Since the
total angular moment $J$ of the halo ($E^*$=3.562 MeV)   state  of $^6$Li is zero, its total orbital moment $L$ and its total spin $S$ are equal ($L$=$S$=0 or 1 ). As the parity is positive, the relative orbital moments $l_{23}$ and
$l_{13}$ take the common value $l$. The  two-dimensional Lagrange
functions have the following form
$$
F_{i_1i_2}(r_{23},r_{13})=  f_{i_1}(r_{13}/h)f_{i_2}(r_{23}/h)/h.
 \eqno(B3)
$$
Here
$$
f_i(x)=(-1)^ix_i^{-1/2}x(x-x_i)^{-1}L_N(x)e^{-x/2} \eqno(B4)
$$
is the one-dimensional Lagrange-Laguerre function, where $L_N(x)$ is
a Laguerre polynomial   and the Laguerre zeros $x_i$ are solution of
$L_N(x_i)$=0 \cite{Bay97,Vin93}. The basis functions
$F_{i_1i_2}(r_{23},r_{13})$ are associated with $N^2$ mesh points
($hx_{i_1},hx_{i_2}$) where they satisfy the Lagrange property
$$
F_{i_1i_2}(hx_{i_1^\prime},hx_{i_2^\prime})\propto\delta_{i_1i_1^\prime}
\delta_{i_2i_2^\prime}.
$$
The  normalization conditions for  the radial wave functions and for
the coefficients $c_{i_1i_2}^{lLS}$ at Gauss  approximation  are
given by Eqs.(36) and (37) of Ref.\cite{Yar02}, respectively. The
scale factor $h$ is a non-linear variational parameter aimed at
adjusting the mesh to the domain of physical interest.

The values of the coefficients $c_{i_1i_2}^{lLS}$ were obtained by D. Baye \cite{Baye2007} for  the both aforesaid  kinds of the $\alpha N$ potential (the M1 and M2 models). Nevertheless one should note only  the following main points. To approximately reproduce the experimental binding energy 0.136 MeV of the second excited state of ${\rm{^6Li}}$,    the potentials  of Refs.\cite{Voron95} and  \cite{Kan79}  were multiplied  by 1.08   and  by 1.01, respectively, as it was also done in  Ref.\cite{Yar02}.   Then, the binding energy was calculated as 0.1365 MeV for the
$\alpha N$ potential Ref.\cite{Voron95} in the M1 model and as
0.1323 MeV for that from Ref.\cite{Kan79} in the M2 model. 
The Coulomb $\alpha p$ interaction was represented by $2e^2 {\rm erf\,}(0.83 r)/r$ 
and took into  account  the finite extension of the $\alpha$ particle. 
The corresponding wave functions contained partial waves $l$=0 to 18
and were obtained with $h=0.3$ and 0.4 fm for the $\alpha N$ potential 
of the M1 model and the M2 model, respectively. 
The calculations were performed with $N$=28, i.e.\ 784 basis states 
per partial wave for a total of 29008 basis states,  for each of the $\alpha N$ potentials. The dominant partial waves for the considered $^6$Li nucleus
were, in decreasing order of importance, ($L,l$)=(0,1), (1,1), 
(0,2), (0,0) and (1,2). The respective probabilities were 
84.6\%,  8.4\%, 3.8\% and 2.2\%  for the M1 model and 
79.1\%, 15.8\%, 3.1\% and 1.6\% for the M2 model \cite{Baye2007}.

\newpage
\newpage
\newpage
\newpage
\newpage

\newpage
\begin{landscape}
\begin{table}[h]
\caption{\label{table1}The three-body asymptotic normalization
functions  $C^{(1)}_{L\,l}$,  $C^{(2)}_{L\,l}$  for the halo ($E^*$=3.562
MeV;\,$J^\pi$=0$^+$;\,T=1)   state of $^6$Li, ratios
$C^{(2)}_{Ll}/C^{(1)}_{Ll}$ and the width of an interval for the
variable $r_{23}$ ($\Delta$) within which the expressions  (\ref{35}) and
(B2) agree within 10\% for different choices of the matching points
($r_{23},r_{13}$) and ($r_{23}+\Delta r_{23},r_{13}+\Delta r_{13}$)
at fixed values of the ratio $r=r_{13}/r_{23}$ for the quantum
numbers $L=$0, 1 and $l$=0, 1 and 2.}
\begin{tabular}{|c|c|c|c|c|c|c|c|c|c|c|c|c|}
\hline
 & & & & &\multicolumn{4}{c|}{M1 model }&\multicolumn{4}{c|}{M2 model}\\
 \cline{6-13}
$(L,l)$&$r$ &$r_{23}$ &$r_{13}$&$\Delta r_{23}$&$\Delta$ &
$C^{(1)}_{Ll}$ & $C^{(2)}_{Ll}$ &
$C^{(2)}_{Ll}/C^{(1)}_{Ll}$&$\Delta$ & $C^{(1)}_{Ll}$ &
$C^{(2)}_{Ll}$ & $C^{(2)}_{Ll}/C^{(1)}_{Ll}$\\
&&fm&fm&fm&fm&  fm$^{-1/4}$ &
fm$^{-1/4}$&&fm&  fm$^{-1/4}$ & fm$^{-1/4}$& \\
\hline 1&2&3&4&5&6&7&8&9&10&11&12&13\\
\hline
(0,0)& 1.1& 14.00& 15.40& 2.000 & 10.0& -130 & -86.7& 0.666& 13.0 &-132  &-84.6 &0.643\\
     &    & 14.25& 15.68&      & 10.2& -125 & -83.7& 0.672& 13.1 &-127  &-82 &0.647\\
     &    & 13.75& 15.13& 2.250 & 10.0& -133 & -87.9& 0.664& 12.8 &-135  &-86.6 &0.639\\
     &    & 14.25& 15.68&      &  10.2& -122 & -82.5& 0.675& 13.3 &-124  &-80.5 &0.659\\
     & 1.3& 10.25& 13.33& 2.000 &  8.9& -127 & -79.8& 0.630& 10.6 &-107  &-67.4 &0.628\\
     &    & 10.75& 13.98&      &  9.1& -116 & -74.4& 0.639& 11.2 &-100  &-63.7 &0.635\\
     &    & 10.25& 13.33& 2.250 &  8.9& -125 & -78.9& 0.631& 11.0 &-105  &-66.0 &0.630\\
     &    & 10.75& 13.98&      &  9.2& -113 & -72.8& 0.642& 11.2 &-100  &-63.7 &0.635\\
     & 2.5&  5.50& 13.75& 2.000 &  5.2& -59.4 & -41.9& 0.706&  6.4 &-53.1  &-37.0 &0.698\\
     &    &  6.00& 15.00&      &  5.4& -53.9 & -38.5& 0.716&  6.4 &-50.3  &-35.4 &0.703\\
     &    &  5.50& 13.75& 2.250 &  5.3& -58.8 & -41.5& 0.706&  6.5 &-51.6  &-36.1 &0.700\\
     &    &  6.00& 15.00&      &  5.5& -52.0 & -37.4& 0.719&  6.5 &-50.0  &-35.1 &0.703\\
(0,1)& 1.1& 16.00& 17.60& 2.000& 6.7 & 127  & 58.0 &0.455 & 8.0  &136  & 61.1& 0.449\\
      &    & 16.50& 18.15&      & 6.6 & 117  & 53.7 &0.459 & 8.2  &124  & 56.2& 0.452\\
      &    & 16.00& 17.60& 2.250& 8.3 & 125  & 57.1 &0.456&9.6    &134  & 60.3& 0.449\\
      &    & 16.50& 18.15&      & 8.4 & 117  & 53.5 &0.459 &10.6  &119  & 54.1& 0.453\\
      &1.5 & 10.25& 15.38& 2.000& 6.2 & 82.4  & 33.0 &0.401 & 6.6  &79.8  & 31.8& 0.398\\
      &    & 10.75& 16.13&      & 6.3 & 77.0  & 31.0 &0.403 & 7.8  &70.7  & 28.4& 0.401\\
      &    & 10.25& 15.38& 2.125& 6.3 & 81.5  & 32.7 &0.401 & 6.7  &78.9  & 31.4& 0.398\\
      &    & 10.75& 16.13&      & 6.4 & 76.6  & 30.9 &0.403 & 8.0  &69.5  & 27.9& 0.402\\
\hline
\end{tabular}
\end{table}
\end{landscape}
\newpage
\begin{landscape}
\begin{table}
\caption{\label{table}Table\,\,1(continue)}
 \vspace{0.3cm}
\begin{tabular}{|c|c|c|c|c|c|c|c|c|c|c|c|c|}
\hline
 & & & & &\multicolumn{4}{c|}{$M1$-model}&\multicolumn{4}{c|}{$M2$-model}\\
 \cline{6-13}
$(L,l)$&$r$ &$r_{23}$ &$r_{13}$&$\Delta r_{23}$&$\Delta$ &
$C^{(1)}_{Ll}$ & $C^{(2)}_{Ll}$ &
$C^{(2)}_{Ll}/C^{(1)}_{Ll}$&$\Delta$ & $C^{(1)}_{Ll}$ &
$C^{(2)}_{Ll}$ & $C^{(2)}_{Ll}/C^{(1)}_{Ll}$\\
&&fm&fm&fm&fm&  fm$^{-1/4}$ &
fm$^{-1/4}$&&fm&  fm$^{-1/4}$ & fm$^{-1/4}$& \\
1&2&3&4&5&6&7&8&9&10&11&12&13\\
\hline
 (0,1)& 2.4&  6.25& 15.00& 2.000& 5.1 & 49.3  & 14.4 &0.292 & 5.6  &48.7  & 14.2& 0.290\\
      &    &  6.75& 16.20&      & 5.1 & 44.4  & 13.0 &0.294 & 5.6  &43.3  & 12.6& 0.292\\
      &    &  6.25& 15.00& 2.125& 5.2 & 48.4  & 14.1 &0.292 & 5.6  &48.4  & 14.1& 0.291\\
      &    &  6.75& 16.20&      & 5.1 & 44.1  & 13.0 &0.294 & 6.4  &42.2  & 12.4& 0.292\\
(0,2)&  1.3  & 17.00& 22.10& 2.000& 7.0 &-11.2  &-3.07 &0.274 &12.3 &-12.7  &-3.43 &0.270 \\
     &       & 17.50& 22.75&      & 6.8 &-11.6  &-3.18 &0.274 &12.5 &-11.6  &-3.16 &0.272\\
     &       & 17.00& 22.10& 2.250& 7.1 &-11.7  &-3.21 &0.273 &12.4 &-11.1  &-3.03 &0.272\\
     &       & 17.50& 22.75&      & 6.8 &-11.9  &-3.25 &0.273 &12.4 &-11.5  &-3.12 &0.272\\
     &  1.9  & 10.00& 19.00& 2.000& 6.6 &-5.06  &-0.895 &0.177 &12.0 &-4.97  &-0.877 &0.177\\
     &       & 10.50& 19.95&      & 6.5 &-4.89  &-0.867 &0.177 &11.7 &-4.76  &-0.841 &0.177\\
     &       & 10.00& 19.00& 2.250& 6.6 &-4.91  &-0.870 &0.177 &11.9 &-5.05  &-0.891 &0.176\\
     &       & 10.50& 19.95&      & 6.5 &-5.11  &-0.905 &0.177 &11.5 &-4.41  &-0.780 &0.177\\
(1,1)& 1.0   & 18.00& 18.00& 2.000& 7.5 &-23.8  &-11.3 &0.477 &10.5 &-27.6  &-13.3 &0.481\\
     &     & 18.50& 18.50&      & 7.5 &-22.2  &-10.6 &0.480 &11.1 &-25.6  &-12.4 &0.484\\
     &     & 18.00& 18.00& 2.250& 7.5 &-23.4  &-11.2 &0.478 &10.7 &-27.1  &-13.1 &0.482\\
     &     & 18.50& 18.50&      & 7.6 &-21.5  &-10.3 &0.481 &11.4 &-25.1  &-12.2 &0.485\\
     & 1.7 & 10.00& 17.00& 2.000& 6.7 &-11.2  &-4.40 &0.393 & 8.0 &-14.3  &-5.59 &0.390\\
     &     & 10.50& 17.85&      & 6.9 &-10.1  &-4.00 &0.398 & 8.5 &-13.0  &-5.12 &0.394\\
     &     & 10.00& 17.00& 2.125& 6.7 &-11.1  &-4.35 &0.393 & 8.1 &-14.2  &-5.53 &0.390\\
     &     & 10.50& 17.85&      & 6.9 &-9.87  &-3.94 &0.399 & 8.6 &-12.9  &-5.07 &0.395\\
(1,2)& 1.3 & 16.50& 21.45& 2.000& 7.1& 1.72  & 0.467& 0.272 &12.1 &2.52   &0.680 &0.270\\
     &     & 17.50& 22.75&      & 6.9& 1.70  & 0.461& 0.272 &12.6 &2.35   &0.636 &0.271\\
     &     & 16.50& 21.45& 2.250& 7.1& 1.71  & 0.466& 0.272 &12.1 &2.49   &0.674 &0.270\\
     &     & 17.50& 22.75&      & 7.0& 1.72  & 0.467& 0.272 &12.6 &2.34   &0.633 &0.271\\
     & 1.9 & 10.00& 19.00& 2.000& 6.8& 1.02  & 0.180& 0.177 &11.5 &1.26   &0.223 &0.177\\
     &     & 12.00& 22.80&      & 6.6& 1.01  & 0.178& 0.177 &11.4 &1.20   &0.212 &0.177\\
     &     & 10.00& 19.00& 2.250& 6.7& 1.000  & 0.178& 0.177 &11.5 &1.25   &0.221 &0.177\\
     &     & 12.00& 22.80&      & 6.5& 0.987  & 0.175& 0.177 &11.5 &1.20   &0.212 &0.177\\
\hline
\end{tabular}
\end{table}
\end{landscape}
\end{document}